\documentclass[12pt]{article}
\usepackage{amsxtra,amssymb,amsmath,enumerate,graphicx,bbm,hyperref,amsthm,color}
\usepackage[active]{srcltx}
\usepackage{t1enc}
\usepackage{aurical}
\usepackage[T1]{fontenc}

\newcommand{\e}{\mathbb{E}}
\newcommand{\E}{\mathbb{E}}

\renewcommand{\P}{\mathbb{P}}
\newcommand{\R}{\mathbb{R}}

\newcommand{\1}{{\mathbf 1}}

\def\ti{{\tau_{i}}}
\def\tip{{\tau_{i+1}}}
\def \Esp#1{{\mathbb E}\left[#1\right]}

\def\be{\begin{align}}
\def\ee{\end{align}}
\def\b*{\begin{eqnarray*}}
\def\e*{\end{eqnarray*}}

\def\vp{\varphi}

\def\be{\begin{eqnarray}}
\def\ee{\end{eqnarray}}
\def\beq{\begin{equation}}
\def\eeq{\end{equation}}
\def\b*{\begin{eqnarray*}}
\def\e*{\end{eqnarray*}}
\def\bi{\begin{itemize}}
\def\ei{\end{itemize}}

\def \1{{\bf 1}}
\def\vp{\varphi}
\def\eps{\varepsilon}

\def\={\;=\;}

\def\x{\times}

\def\Esp#1{\mathbb{E}\left[#1\right]}

\def \proof{{\noindent \bf Proof. }}
\def \ep{\hbox{ }\hfill$\Box$}
 \def\reff#1{{\rm(\ref{#1})}}
 \def\And{\;\mbox{ and }\;}

 \def\vs#1{\vspace{#1mm}}

\def\ti{{t_i}}
\def\tip{ {t_{i+1}} }
\def\tim{ {t_{i-1}}}

\def \D{\mathbb{D}}
\def \E{\mathbb{E}}
\def \F{\mathbb{F}}
\def \G{\mathbb{G}}

\def \P{\mathbb{P}}

\def \R{\mathbb{R}}

\def\Ac{{\cal A}}

\def\Ec{{\cal E}}
\def\Fc{{\cal F}}
\def\Gc{{\cal G}}

\def\Mc{{\cal M}}

\def\Oc{{\cal O}}

\def\Uc{{\cal U}}

\def\Lb{{\mathbf L}}

\newtheorem{Theorem}{Theorem}[part]

\newtheorem{Proposition}{Proposition}[part]

\newtheorem{Lemma}{Lemma}[part]
\newtheorem{Corollary}{Corollary}[part]
\newtheorem{Remark}{Remark}[part]

\makeatletter \@addtoreset{equation}{section}

\@addtoreset{Definition}{section}

\@addtoreset{Theorem}{section}

\@addtoreset{Proposition}{section}

\@addtoreset{Assumption}{section}

\@addtoreset{Corollary}{section}

\@addtoreset{Lemma}{section}

\@addtoreset{Remark}{section}

\@addtoreset{Example}{section}

  \def\ti{t_{i}^{n}}
\def\tip{t_{i+1}^{n}}
\def\tim{t_{i-1}^{n}}

\def\Yn{Y^{n }}
\def\deltan{\delta^{n }}
\def\Vn{V^{n }}
\def\Sn{X^{n }}

\def\Ve{V^{\eps }}
\def\Se{X^{\eps }}
\def\Ye{Y^{\eps }}

\def\xrm{ {\rm{x}}}
\def\I{{\Delta \xrm}}
\def\J{{\mathfrak  I}}
\def\X{\R}

\def\D{{\rm D}}
\def\G{{\rm G}}
\def\w{\hat w}
\def\Mb{\mathbf{M}}

\begin{document}

\title{Almost-sure hedging with permanent price impact}

\author{B. Bouchard\thanks{CEREMADE, Universit\'e Paris Dauphine and CREST-ENSAE. Research supported by ANR Liquirisk and Investissements d'Avenir (ANR-11-IDEX-0003/Labex Ecodec/ANR-11-LABX-0047).
     } \And G. Loeper\thanks{BNP-Paribas and FiQuant - Chaire de finance quantitative} \And Y. Zou \thanks{CEREMADE, Universit\'e Paris Dauphine and CREST-ENSAE).
     }}
\maketitle

\begin{abstract} We consider a financial model with permanent price impact. Continuous time trading dynamics are derived as the limit of discrete rebalancing policies. We then study the problem of super-hedging a European option. Our main result is the derivation of a quasi-linear pricing equation. It holds in the sense of viscosity solutions. When it admits a smooth solution, it provides a perfect hedging strategy.
\end{abstract}


\vspace{1em}

\noindent \emph{AMS 2010 Subject Classification:}
%%60H05 %Stochastic integrals
91B28; %Finance, portfolios, investment
93E20; %Optimal stochastic control
49L20 %Dynamic programming method
%%60G44, %Martingales with continuous parameter
%%91B30  %Risk theory, insurance

\section*{Introduction}
Two of the fundamental assumptions in the Black and Scholes approach for option hedging are that the price dynamics are unaffected by the hedger's behaviour, and that he can trade unrestricted amounts of asset at the instantaneous value of the price process. In other words, it relies on the absence of market impact and of liquidity costs or liquidity constraints. This work addresses the problem of option hedging under a price dynamics model that incorporates directly the hedger's trading activity, and hence that violates those two assumptions. 

In the literature, one finds numerous studies related to this  topic. Some of them incorporate liquidity costs but no price impact, the price curve is not affected by the trading strategy. In the setting of \cite{cetin2004liquidity}, this does not affect the super-hedging price because trading can essentially be done in a bounded variation manner at the marginal spot price at the origine of the curve. However, if additional restrictions are imposed on admissible strategies, this  leads to  a modified pricing equation, which exhibits a quadratic term in the second order derivative of the solution, and renders the pricing equation fully non-linear, and even not unconditionally parabolic, see \cite{CetinSonerTouzi} and  \cite{SonTouzDyn}. Another branch of literature focuses on the derivation of the price dynamics through clearing condition. In the  papers \cite{Frey}, \cite{Sircar}, \cite{Schon}, the authors work on supply and demand curves that arise from ``reference'' and ``program'' traders (i.e. option hedgers) to establish a modified price dynamics, but do not take into account the liquidity costs, see also \cite{Liu}. This approach also leads to non-linear pde's, but the non-linearity comes from a modified volatility process rather than from a liquidity cost source term. Finally,    the series of papers \cite{SonTouz}, \cite{SonTouz2}, \cite{CheriSonTouz} address the liquidity issue indirectly by imposing  bounds on the ``gamma'' of admissible trading strategies, no liquidity cost or price impact are modeled explicitly. 

More recently, \cite{Loeper} and \cite{AbergelLoeper} have considered a novel approach in which the price dynamic is driven by the sum of a classical Wiener process and a (locally) linear market impact term. The linear market impact mechanism induces a modified volatility process, as well as a non trivial average execution price. However, the trader starts his hedging with the correct position in stocks and does not have to unwind his final position (this corresponds to ``covered'' options with delivery).  Those combined effects lead to a fully non-linear pde giving the exact replication strategy, which is not always parabolic depending on the ratio between the instantaneous market impact (liquidity costs) and permanent market impact. 

In this paper we build on the same framework as \cite{Loeper}, in the case where the instantaneous market impact equals the permanent impact (no relaxation effect), and
go one step further by considering the effect of (possibly) unwinding the portfolio at maturity, and of building the initial portfolio. Consequently the spot ``jumps'' at initial time when building the hedge portfolio, and at maturity when unwinding it (depending on the nature of the payoff - delivery can also be made in stocks). In this framework, we find that the optimal super-replication strategy follows a modified quasi-linear Black and Scholes pde. Although the underlying model is similar to the one proposed by the second author \cite{Loeper}, the pricing pde is therefore fundamentally different   (quasi-linear vs fully non-linear).
% because of allowing the final liquidation of the portfolio at maturity. More precisely, in this work, the space of admissible hedging strategies allow the option seller to execute a large order at maturity, which can impact instantaneously the value of the final claim, this was not allowed in \cite{Loeper} and \cite{AbergelLoeper}. 

Concerning the mathematical approach, while in \cite{Loeper} the author focused on exhibiting an exact replication strategy by a verification approach, in this work we follow a stochastic target approach and derive the pde from a dynamic programming principle. The difficulty is that, because of  the market impact mechanism, the state process must be described by the asset price and the hedger's portfolio (i.e.~the amount of risky asset detained by the hedger) and this leads to a highly singular control problem. It is overcome  by a suitable change of variable which allows one to reduce to a zero initial position in the risky asset and state a version of the geometric dynamic programming principle in terms of the post-portfolio liquidation asset price process: the price that would be obtained if the trader was liquidating his position immediately.  

The paper is organized as follows. In Section \ref{sec: dynamics}, we present the impact rule and derive continuous time trading dynamics as limits of discrete time rebalancing policies. The super-hedging problem is set in Section \ref{sec: super hedging} as a stochastic target problem. We first prove a suitable version of the geometric dynamic programming and then derive the corresponding pde in the viscosity solution sense. Uniqueness and regularity are established under suitable assumptions. We finally  further discuss the case of a constant impact coefficients, to provide a better understanding of the ``hedging strategy''.  

\vspace{5mm}

\noindent{\bf General notations.}
Given a function $\phi$, we denote by  $\phi'$ and $\phi^{''}$ its first and second order derivatives if they exist. When $\phi$ depends on several arguments, we use the notations $\partial_{x}\phi$, $\partial^{2}_{xx} \phi$ to denote the first and second order partial derivatives with respect to its $x$-argument, and write   $\partial_{xy}^{2}\phi$ for the cross second order derivative  in its $(x,y)$-argument.

All over this paper,  $\Omega$ is the canonical space of continuous functions on $\R_{+}$ starting at $0$, $\P$ is the Wiener measure, $W$ is the canonical process, and $\F=(\Fc_{t})_{t\ge 0}$ is its augmented raw filtration.  All random variables are defined on $(\Omega,\Fc_{\infty},\P)$.
$\Lb_{0}$ (resp. $\Lb_{2}$) denotes the space of (resp. square integrable) $\R^{n}$-valued random variables, while $\Lb^{\lambda}_{0}$ (resp. $\Lb^{\lambda}_{2}$) stands for the collection of predictable $\R^{n}$-valued processes $\vartheta$ (resp. such that   $\|\vartheta\|_{\Lb^{\lambda}_{2}}:=\E[\int_{0}^{\infty}|\vartheta_{s}|^{2}ds]^{\frac12}$). The integer  $n\ge 1$ is given by the context and $|x|$ denote the Euclidean norm of $x\in \R^{n}$.

Given a stochastic process $\xi$, we shall always denote by $\xi^{c}$ its continuous part.

%%%%%%%%%%%%%%%%%%%%%%%%%%%%%%
\section{Portfolio and price dynamics}\label{sec: dynamics}

This section is devoted to the derivation of our model with  continuous time trading. We first consider the situation where a trading signal is given by a continuous It\^{o} process and the position in stock is rebalanced in discrete time. In this case, the dynamics of the stock price and the wealth process are given according to our impact rule. A first continuous time trading dynamic  is obtained by letting the time between two consecutive trades vanish. Then, we  incorporate jumps as the limit of continuous trading on a short time horizon.

We restrict here to a single stock market. This is only for   simplicity, the extension to a multi-dimensional market is just a matter of notations.

\subsection{Impact rules}

We model the impact of a strategy on the price process through an impact function $f$: 
 the price variation du to buying a (infinitesimal)   number $\delta\in \R$ of shares is $\delta f(x)$,  if the price of the asset is $x$ before the trade. The cost of buying the additional  $\delta$ units is given by
$$
\delta x+\frac12 \delta^{2} f(x)=\delta \int_{0}^{\delta}\frac1\delta (x+f(x) \iota) d\iota,
$$
in which
$$
 \int_{0}^{\delta}\frac1\delta (x+f(x) \iota) d\iota
$$
should be interpreted as the average cost for each additional unit. Between two times of trading $\tau_{1}\le \tau_{2}$, the dynamics of the stock is given by the strong solution of the stochastic differential equation
$$
dX_{t}=\mu(X_{t})dt +\sigma(X_{t}) dW_{t}.
$$

All over this paper, we assume that
\b*
\begin{array}{c}
\mbox{$f\in C^{2}_{b}$   and is (strictly) positive,} \\
\mbox{ $(\mu,\sigma,\sigma^{-1})$ is Lipschitz and bounded. }
  \end{array}
  &{\bf \rm (H1)}
\e*

\begin{Remark}
a. We restrict here to an  impact rule which is linear in   the size of the order. However, note that  in the following it will only be applied to order of infinitesimal size (at the limit).  One would therefore obtain the same final dynamics \reff{eq: S lim conti avec jump}-\reff{eq: V lim conti avec jump} below by considering a more general impact rule $\delta \mapsto F(x,\delta)$ whenever is satisfies $F(x,0){=\partial^{2}_{\delta \delta}F(x,0)}=0$ and  $\partial_{\delta}F(x,0)=f(x)$. See Remark \ref{rem: F general} below. Otherwise stated, for our analysis, we only need to consider the value and the slope at $\delta=0$ of the impact function.  

b. A typical example of such a function is  $F=\I$ where
\be\label{eq: def I}
\I(x,\delta):=\xrm(x,\delta)-x\,,\,
\ee
with  $\xrm(x,\cdot)$ defined as  the solution of
\be\label{eq: def xrm}
\xrm(x,\cdot)&=&x+\int_{0}^{\cdot} f(\xrm(x,s))ds.
\ee
The curve $\xrm$ has a natural interpretation. For an order of small size $\Delta\iota$, the stock price  jumps from $x$ to $x+\Delta\iota f(x)\simeq\xrm(x,\Delta\iota)$. Passing another order of size $\Delta\iota$ makes it move again to approximately $\xrm(\xrm(x,\Delta\iota),\Delta\iota)=\xrm(x,2\Delta \iota)$, etc. Passing to the limit $\Delta\iota\to 0$ but keeping the total trade size equal to $\delta$ provides asymptotically a price move equal to $\I(x,\delta)$.

This specific curve will play a central role in our analysis, see Section \ref{subsec: add jump}.
\end{Remark}

%%%%%%%%%%%%%%%%%%%%%%%%%%
\subsection{Discrete rebalancing  from a continuous signal and continuous time trading  limit}

We first consider the situation in which the number of shares the  trader would like to hold is given by a continuous It\^{o} process  $Y$ of the form
\be\label{eq: def Y sans saut}
Y =Y_{0}+\int_{0}^{\cdot} b_{s} ds +\int_{0}^{\cdot}a_{s}dW_{s},
\ee
where 
\b*
&(a,b)  \in \Ac:=\cup_{k}\Ac_{k},&\\
&\Ac_{k}:=\{(a,b)\in \Lb_{0}^{\lambda} : |(a,b)|\le k\;dt\x d\P-{\rm a.e.}\} \mbox{ for } k>0. &
\e*
\vs2

In order to derive our continuous time trading dynamics, we consider the corresponding discrete time rebalancing policy set on a  time grid
$$
\ti:=iT/n,\;i=0,\ldots,n, \;n\ge 1,
$$
and then pass to the limit $n\to \infty$.

If the trader  only changes the composition of his portfolio at the discrete times $\ti$, then he holds $Y_{\ti}$ stocks on each time interval $[\ti,\tip)$. Otherwise stated, the number of shares actually held at $t\le T$ is
\be\label{eq: def Yn sans saut}
\Yn_{t}:=\sum_{i=0}^{n-1} Y_{\ti}\1_{\{\ti\le t<\tip\}}+Y_{T}\1_{\{t=T\}}
\ee
and the number of purchased shares  is
$$
\deltan_{t}:=\sum_{i=1}^{n}\1_{\{t=\ti\}} (Y_{\ti}-Y_{\tim}). 
$$
Given our impact rule, the corresponding dynamics for the stock price  process   is
\be\label{eq: dyna Sn}
\Sn  = X_{0} + \int_{0}^{\cdot}\mu(\Sn_{s})ds +\int_{0}^{\cdot} \sigma(\Sn_{s}) dW_{s} + \sum_{i=1}^{n}\1_{[\ti,T]} \deltan_{\ti} f(\Sn_{\ti-}),
\ee
in which $X_{0}$ is a constant.

To describe the portfolio process, we provide the dynamics of the sum $\Vn$ of the amount of cash held and the potential amount $\Yn \Sn$ associated to the position in stocks:
\be\label{eq: def Vn}
V^{n}=\mbox{cash position } + \Yn \Sn.
\ee
Observe that this is not the liquidation value of the portfolio, except when $\Yn=0$, as the liquidation of $\Yn$ stocks will have an impact on the market and does not generate a gain equal to $\Yn \Sn$.
However, if we keep $\Yn$ in mind, the couple $(\Vn,\Yn)$ gives the exact composition in cash and stocks of the portfolio. By a slight abuse of language, we call $\Vn$ the portfolio value or wealth process.

Assuming that the risk free rate is zero (for  {ease of notations}), its dynamics is given by
\begin{equation}
\Vn = V_{0 } + \int_{0}^{\cdot}\Yn_{s-}d\Sn_{s} + \sum_{i=1}^{n}\1_{[\ti,T]}  \frac12 ( \deltan_{\ti})^{2}f(\Sn_{\ti-}), \label{eq: dyna Vn}
\end{equation}
or equivalently
\begin{align}
\Vn &= V_{0 } +\sum_{i=1}^{n}  \1_{[\tim,T]} Y_{\tim} (\Sn_{\cdot \wedge \ti-}-\Sn_{\tim})  \nonumber\\
&\; + \sum_{i=1}^{n}   \1_{[\ti,T]} \left[   \frac12 ( \deltan_{\ti})^{2}f(\Sn_{\ti-})  + Y_{\tim} \deltan_{\ti} f(\Sn_{\ti-} )\right],  \label{eq: dyna Vn developpee}
\end{align}
in  which $V_{0 }\in \R$. Let us comment this formula. The first term on the right-hand side corresponds to the evolution of the portfolio value strictly between two trades ; it is given by the number of shares held multiplied by the price increment. When a trade of size $\deltan_{\ti}$ occurs at time $\ti$, the costs of buying the stocks is $  2^{-1} ( \deltan_{\ti})^{2}f(\Sn_{\ti-}) + \deltan_{\ti}\Sn_{\ti-}$ but it provides $\deltan_{\ti}$ more stocks, on top of the $\Yn_{\ti-}=Y_{\tim}$ units that are already in the portfolio. After the price's move generated by the trade, the stocks are evaluated at  $\Sn_{\ti}$. The increment in value du to the price's move and the additional position is therefore $\deltan_{\ti}\Sn_{\ti} $ $+$ $\Yn_{\ti-}(\Sn_{\ti} -\Sn_{\ti-} ) $. Since $\Sn_{\ti} -\Sn_{\ti-}$ $=$ $\deltan_{\ti} f(\Sn_{\ti-}) $, we obtain \reff{eq: dyna Vn developpee}, a compact version of which is given in   \reff{eq: dyna Vn}.

\vs2

Our continuous time trading dynamics are obtained by passing to the limit $n\to \infty$, i.e.~by considering faster and faster rebalancing strategies.

\begin{Proposition}\label{prop : lim quand n to infty} Let $Z:=(X,Y,V)$ where $Y$ is defined as in \reff{eq: def Y sans saut} for some $(a,b)\in \Ac$, and  $(X,V)$ solves
\begin{equation}
X=X_{0 }+\int_{0}^{\cdot} \sigma(X_{s}) dW_{s} + \int_{0}^{\cdot} f(X_{s}) dY_{s}+\int_{0}^{\cdot} (\mu(X_{s})+a_{s}(\sigma f')(X_{s}) )ds  \label{eq: S lim conti}
\end{equation}
and
\begin{equation}
V=V_{0 }+ \int_{0}^{\cdot} Y_{s}dX_{s}+\frac12 \int_{0}^{\cdot} a^{2}_{s} f(X_{s}) ds.\label{eq: V lim conti}
\end{equation}
Let $Z^{n}:=(X^{n},Y^{n},V^{n})$ be defined as in \reff{eq: dyna Sn}-\reff{eq: def Yn sans saut}-\reff{eq: dyna Vn}.
Then,  there exists a constant $C>0$ such that
\b*
\sup_{[0,T]}\Esp{  |Z^{n} -Z |^{2} }\le Cn^{-1}
\e*
for all $n\ge 1$.
 \end{Proposition}

\proof This follows standard arguments and we only provide the main ideas. In all this proof, we denote by $C$ a generic positive constant which does not depend on $n$ nor $i\le n$, and may change from line to line.    We shall use repeatedly  {\rm (H1)} and the fact that $a$ and $b$ are bounded by some constant $k$, in the $dt\x d\P$-a.e.~sense.

 {a.}  The convergence of the process $\Yn$ is obvious:
 \be\label{eq: estime Y-Yn Lp}
\sup_{[0,T]} \Esp{ |\Yn -Y |^{2}  }\le C n^{-1}.
 \ee
For later use, set $\Delta X^{n}:=X-X^{n}$ and also observe that the estimate 
\begin{equation} 
\sup_{[\tim,\ti)}\Esp{| \Delta X^{n}|^{2}} \le  \Esp{| \Delta X^{n}_{\tim} |^{2}}(1+Cn^{-1})+ Cn^{-1},\label{eq: ito sur delta X p}
\end{equation}
is standard. 
We now set 
$$
\tilde X^{n}_{t}:=X^{n}_{t}+A^{n}_{t}+ B^{n}_{t}, \;\;\tim\le t\le \ti,
$$
where 
\b*
A^{n}_{t}&:=& \int_{\tim}^{t}  f(\Sn_{s})dY_{s} + \int_{\tim}^{t} a_{s }(\sigma f')(\Sn_{s})ds \\
B^{n}_{t}&:=&\int_{\tim}^{t} (Y_{s}-Y_{\tim}) (\mu f'+ \frac12  \sigma^2  f^{''} ) (\Sn_{s})ds+\int_{\tim}^{t} (Y_{s}-Y_{\tim}) (\sigma f')(\Sn_{s}) dW_{s}.
\e*
Since $A^{n}_{\ti}+B^{n}_{\ti}=\deltan_{\ti} f(\Sn_{\ti-})$, we have $$\tilde X^{n}_{\ti}=X^{n}_{\ti}.$$ 
Set $\Delta \tilde X^{n}:= X-\tilde X^{n}${, 
$\beta^{1}:= bf +a \sigma f'$ and $\beta^{2}:=af$},
so that 
\b*
d|\Delta \tilde X^{n}_{t}|^{2}&=&2\Delta \tilde X^{n}_{t}  [(\mu+\beta^{1}_{t})(X_{t})-(\mu+\beta^{1}_{t})(\Sn_{t})]dt \\
&&+\;[ (\sigma+\beta^{2}_{t})(X_{t})-(\sigma+\beta^{2}_{t})(\Sn_{t})- (Y_{t}-Y_{\tim}) (\sigma f')(\Sn_{t}) ]^{2} dt \\
&&-\;2\Delta \tilde X^{n}_{t}(Y_{t}-Y_{\tim}) (\mu f'+ \frac12  \sigma^2  f^{''} ) (\Sn_{t})dt\\
&&+\;2\Delta \tilde X^{n}_{t} [ (\sigma+\beta^{2}_{t})(X_{t})-(\sigma+\beta^{2}_{t})(\Sn_{t})]dW_{t}\\
&&-\;2\Delta \tilde X^{n}_{t}  (Y_{t}-Y_{\tim}) (\sigma f')(\Sn_{t}) dW_{t}.
\e*
In view of \reff{eq: estime Y-Yn Lp}-\reff{eq: ito sur delta X p}, this implies 
\begin{align*}
\Esp{|\Delta \tilde X^{n}_{t}|^{2}}&\le \Esp{|\Delta   X^{n}_{\tim}|^{2}} +
C\Esp{\int_{\tim}^{t} (|\Delta   \tilde X^{n}_{s}|^{2}+|X_{s}-X^{n}_{s}|^{2}+|Y_{s}-Y_{\tim}|^{2})ds }\\
&\le \Esp{|\Delta   X^{n}_{\tim}|^{2}}(1+Cn^{-1}) +
C\Esp{\int_{\tim}^{t} |\Delta \tilde X^{n}_{s}|^{2}ds +n^{-2}},
\end{align*}
and therefore 
\be 
\sup_{[\tim,\ti]}\Esp{| \Delta \tilde X^{n} |^{2}}& \le &   \Esp{| \Delta X^{n}_{\tim} |^{2} } (1+Cn^{-1}) +Cn^{-2}, \label{eq: ito sur delta tilde X p}
\ee
by Gronwall's Lemma.  Since $\tilde X^{n}_{\ti}=X^{n}_{\ti}$,  this shows that  
$$
\Esp{| \Delta X^{n}_{\ti} |^{2}}\le Cn^{-1}\;\;\mbox{ for all } i\le n. 
$$
Plugging this inequality in \reff{eq: ito sur delta X p}, we then deduce  
\be \label{eq: S - Sn en Lp}
\sup_{[\tim,\ti]}\Esp{| \Delta X^{n}  |^{2}}
&\le &  Cn^{-1}  \;\;\mbox{ for all } i\le n. 
\ee

 b.  We now consider the difference $V-\Vn$.  It follows from \reff{eq: dyna Vn developpee} that 
 \begin{align*}
 \Vn_{\ti}&=\Vn_{\tim}+\int_{\tim}^{\ti} Y_{\tim}\mu(X^{n}_{s}) ds + \int_{\tim}^{\ti} Y_{\tim}\sigma(X^{n}_{s}) dW_{s}\\
 &+ \int_{\tim}^{\ti}\left( \frac12 a^{2}_{s}f(X^{n}_{s})+Y_{\tim}a_{s}(f'\sigma)(X^{n}_{s})\right)ds+ \int_{\tim}^{\ti}Y_{\tim} f(X^{n}_{s})dY_{s}\\
  &+\int_{\tim}^{\ti} \alpha^{1n}_{s} ds +\int_{\tim}^{\ti}\alpha^{2n}_{s}dW_{s}
 \end{align*}
 where, by    \reff{eq: estime Y-Yn Lp},
  $\alpha^{1n}$ and $\alpha^{2n}$ are adapted processes satisfying
\b*
  \sup_{[\tim,\ti)}\E[|\alpha^{1n}|^{2}+|\alpha^{2n}|^{2}] \le C n^{-1}. 
 \e*
In view of  \reff{eq: estime Y-Yn Lp}-\reff{eq: S - Sn en Lp}, this leads {to}
  \be\label{eq: Vn en fonction betas}
  \Vn_{\ti}=\Vn_{\tim}+ V_{\ti}-V_{\tim} +  \int_{\tim}^{\ti} \gamma^{1n}_{s} ds +  \int_{\tim}^{\ti} \gamma^{2n}_{s} dW_{s} 
  \ee
where   
  $\gamma^{1n}$ and $\gamma^{2n}$ are adapted processes satisfying
\be\label{eq: estime beta1n beta2n} 
  \sup_{[\tim,\ti)}\E[|\gamma^{1n}|^{2}+|\gamma^{2n}|^{2}] \le C n^{-1}. 
 \ee
  Set  
 $$
  \tilde V^{n}_{t}:=\Vn_{\tim}+V_{t}-V_{{\tim}}  +  \int_{\tim}^{t} \gamma^{1n}_{s} ds +  \int_{\tim}^{t} \gamma^{2n}_{s} dW_{s},\;\;\tim\le t \le \ti.
  $$
  Then, by applying It\^{o}'s Lemma to $|\tilde V^{n}_{t}-V_{t}|^{2}$, using \reff{eq: estime beta1n beta2n} and Gronwall's Lemma, we obtain 
  $$
 \sup_{[\tim,\ti]}  \Esp{|\tilde V^{n}-V|^{2}}\le  \Esp{|  V^{n}_{\tim}-V_{\tim}|^{2}}(1+Cn^{-1}) + Cn^{-2},
  $$
  so that, by  the identity $\tilde V^{n}_{\ti}=V^{n}_{\ti}$ and an induction,  
  $$
  \Esp{|\Vn_{\ti}-V_{\ti}|^{2}}\le C  n^{-1},\;\;i\le n. 
  $$
We conclude by observing that 
  \b*
 \Esp{|  V^{n}_{t}-V_{t}|^{2}}&\le& C\Esp{|  V^{n}_{\tim}-V_{\tim}|^{2}+ |  V^{n}_{\tim}-V^{n}_{t}|^{2}+|  V_{\tim}-V_{t}|^{2}}
 \\
 &\le& C\left( \Esp{|  V^{n}_{\tim}-V_{\tim}|^{2} }+n^{-1}\right),
  \e*
  for $\tim \le t<\ti$.
\ep

\begin{Remark}\label{rem: F general} If the impact function $\delta f(x)$ was replaced by a more general $C^{2}_{b}$ one of the form $F(x,\delta)$, with $F(x,0)=$ $\partial^{2}_{\delta\delta}F(x,0)=0$, the computations made in the above proof would only lead to   terms of the from $\partial_{\delta}F(X,0)dY$ and $a\sigma(X) \partial^{2}_{x \delta }F(X,0)$ in place of $f(X)dY$ and $a(\sigma f')(X)$ in the dynamics \reff{eq: S lim conti}. Similarly, the term $a^{2}f(X)$ would be replaced by $a^{2}\partial_{\delta}F(X,0)$ in \reff{eq: V lim conti}.
\end{Remark}

\subsection{Jumps and large orders splitting}\label{subsec: add jump}

We now explain how we incorporate jumps in our dynamics. Let $\Uc_{k}$ denote the set of random $\{0,\cdots,k\}$-valued measures $\nu$ supported by $[-k,k]\x[0,T]$ that are adapted in the sense that
$t\mapsto \nu(A\x [0,t])$ is adapted for all Borel subset $A$ of  $[-k,k]$. We set $$\Uc:=\cup_{k\ge 0}\ \Uc_{k}.$$ Note that an element $\nu$ of $\Uc$ can be written in the form
\be\label{eq: nu en terme de somme}
\nu(A,[0,t])=\sum_{j=1}^{k} \1_{\{(\delta_{j},\tau_{j})\in A\x [0,t]\}}
\ee
in which $0\le \tau_{1}< \cdots< \tau_{k}\le T$ are  stopping times and each $\delta_{j}$ is a real-valued $\Fc_{\tau_{j}}$-random variable.

Then, given $(a,b,\nu)\in \Ac\x \Uc$, we define the trading signal as
\be\label{eq: def Y avec saut simple}
Y =Y_{0-}+\int_{0}^{\cdot} b_{s} ds +\int_{0}^{\cdot}a_{s}dW_{s}+\int_{0}^{\cdot} \int \delta \nu(d\delta,ds),
\ee
where  $Y_{0-}\in \R$.

 In view of the previous sections, we assume that the dynamics of the stock price and portfolio value processes are given by \reff{eq: S lim conti}-\reff{eq: V lim conti} when  $Y$ has no jump. We incorporate jumps by assuming that the trader    follows the natural idea of splitting a large order $\delta_{j}$ into small pieces on a small time interval. This is a current practice which aims at avoiding having a too large impact, and paying a too high liquidity cost. Given the asymptotic already derived in the previous section,   we can reduce to the case where this is done continuously at a constant rate $\delta_{j}/\eps$ on  $[\tau_{j},\tau_{j}+\eps]$, for some $\eps>0$.  We denote by $(X_{0-},V_{0-})$ the initial   price and portfolio values.
Then, the number of stocks in the portfolio associated to a strategy $(a,b,\nu)\in \Ac_{k}\x\Uc_{k}$ is   given by
\begin{equation}
 \Ye= Y+\sum_{j=1}^{k}\1_{[\tau_{j},T]} \left[-\delta_{j}+  \eps^{-1} \delta_{j} (\cdot \wedge (\tau_{j}+\eps)-\tau_{j})\right], \label{eq: Yne jumps}
\end{equation}
and the corresponding   stock price and  portfolio value dynamics are
\begin{align}
\Se&=X_{0-}+\int_{0}^{\cdot} \sigma(\Se_{s}) dW_{s} + \int_{0}^{\cdot} f(\Se_{s}) d\Ye_{s}+\int_{0}^{\cdot} (\mu(\Se_{s})+a_{s}(\sigma f')(\Se_{s}) )ds \label{eq: Sne jumps}\\
\Ve&=V_{0-}+ \int_{0}^{\cdot} \Ye_{s}d\Se_{s}+\frac12 \int_{0}^{\cdot} a^{2}_{s} f(\Se_{s}) ds .    \label{eq: Vne jumps}
\end{align}

When passing to the limit  $\eps\to 0$, we obtain the  convergence of $Z^{\eps}:=(X^{\eps},Y^{\eps},V^{\eps})$ to $Z=(X,Y,V)$ with $(X,V)$ defined in  \reff{eq: S lim conti avec jump}-\reff{eq: V lim conti avec jump} below. In the following, we only state the convergence of the terminal values, see the proof for a more complete description. It uses the curve $\xrm$ defined in \reff{eq: def xrm} above, recall also \reff{eq: def I}.

\begin{Proposition}  Given $(a,b,\nu)\in \Ac\x\Uc$, let  $Z=(X,Y,V)$ be defined by \reff{eq: def Y avec saut simple} and
\begin{eqnarray}
X&=&X_{0-}+\int_{0}^{\cdot} \sigma(X_{s}) dW_{s} + \int_{0}^{\cdot} f(X_{s}) dY^{c}_{s}+\int_{0}^{\cdot} (\mu(X_{s})+a_{s}(\sigma f')(X_{s}) )ds \nonumber\\
&&+\;\int_{0}^{\cdot} \int \I(X_{s-},\delta)\nu(d\delta,ds)\label{eq: S lim conti avec jump}\\
V&=&V_{0-}+ \int_{0}^{\cdot} Y_{s}dX^{c}_{s}+\frac12 \int_{0}^{\cdot} a^{2}_{s} f(X_{s}) ds\nonumber\\
&&+\;\int_{0}^{\cdot} \int\left(Y_{s-}\I(X_{s-},\delta )+ \J(X_{s-},\delta )\right)\nu(d\delta,ds)\, \label{eq: V lim conti avec jump}
\end{eqnarray}
where
\be\label{eq: def J}
\J(x,z ):= \int_{0}^{z} s f(\xrm(x,s))ds,\;\,\mbox{ for }x,z\in \R.
\ee
Set  $Z^{\eps}:=(X^{\eps},V^{\eps},Y^{\eps})$. Then,  there exists a constant $C>0$ such that
\begin{equation*}\label{eq: estimation Ve-V Se-S}
 \Esp{  |Z^{\eps}_{ T+\eps}-Z_{  T}|^{2}  }\le C(\eps +\P[\sup_{t\le T} \nu(\R,[t,t+\eps])\ge 2]^{\frac12}),
\end{equation*}
for all $\eps\in (0,1)$. Moreover, 
$$
\lim_{\eps\to 0 }\P[\sup_{t\le T} \nu(\R,[t,t+\eps])\ge 2]=0.
$$
\end{Proposition}

\proof  In all this proof, we denote by $C$ a generic positive constant which does not depend on $\eps$, and may change from line to line.  Here again, we shall use repeatedly  {\rm (H1)} and the fact that $a$ and $b$ are bounded by some constant $k$, in the $dt\x d\P$-a.e.~sense. 

Let $\nu$ be of the form \reff{eq: nu en terme de somme} for some $k\ge 0$ and note that the last claim simply follows from the fact that $\{\tau_{j+1}-\tau_{j}\ge \eps\}\uparrow$ $\Omega$ up to a $\P$-null set for all $j\le k$.

\noindent Step 1.  We first consider the case where $\tau_{j+1}\ge \tau_{j}+\eps$ for all $j\ge 1$.
Again, the estimate  on $|Z^{\eps}_{ T+\eps}-Z_{  T}|$ follows from simple observations and  standard estimates, and we only highlight the main ideas.
We will indeed prove that for $1\le j\le k+1$ 
\begin{equation}\label{eq: estimated Zeps on stopping times intervals}
 \Esp{\sup_{ [\tau_{j-1}+\eps,\tau_{j})} |Z -Z^{\eps} |^{2}+\sup_{0\le s\le \eps}\E[  |Z_{\tau_{j}+s}-Z^{\eps}_{\tau_{j}+\eps}|^{2}} \le C\eps,
\end{equation}
where  we use the convention $\tau_{0}=0$ and  $\tau_{k+1}=T$.
The result is trivial for $(\Ye,Y)$ since they are equal on each intervalle $[\tau_{j-1}+\eps,\tau_{j})$ and $(a,b)$ is bounded.

a. We first prove a stronger result for $(\Se,X)$. Fix  $p\in \{2,4\}$.    Let $\xrm^{\eps}$ be the solution of the ordinary differential equation
$$
\xrm^{\eps}_{t}=X_{\tau_{j}-}+\int_{0}^{t}\frac{\delta_{j}}{\eps}f(\xrm^{\eps}_{s})ds.
$$
Set $\Delta  X^{\eps}:=X^{\eps}-\xrm^{\eps}_{\cdot-\tau_{j}}$. It\^{o}'s Lemma leads to 
\begin{align*}
d(\Delta  X^{\eps}_{t})^{p}&=p(\Delta  X^{\eps}_{t})^{p-1}\alpha^{1,\eps}_{t} dt +\frac{p(p-1)}{2} (\Delta  X^{\eps}_{t})^{p-2}(\alpha^{2,\eps}_{t})^{2} dt\\
&+  p(\Delta  X^{\eps}_{t})^{p-1}\alpha^{2,\eps}_{t} dW_{t}\\
&+p \frac{\delta_{j}}{\eps}(\Delta  X^{\eps}_{t})^{p-1}(f(X^{\eps}_{t})-f(\xrm^{\eps}_{t-\tau_{j}}))dt
\end{align*} 
on $[\tau_{j},\tau_{j}+\eps]$, in which $\alpha^{1,\eps}$ and $\alpha^{2,\eps}$ are bounded processes.
The inequality $x^{p-1}\le x^{p-2}+x^{p}$, the Lipschitz continuity of $f$ and  Gronwall's Lemma then imply
$$
\sup_{0\le t\le \eps }\E\left[|\Se_{\tau_{j}+t}- \xrm^{\eps}_{t}|^{p}\right]\le C \Esp{|\Se_{\tau_{j}-}-X_{\tau_{j}-}|^{p}+\int_{0}^{\eps} |\Se_{\tau_{j}+s}- \xrm^{\eps}_{s}|^{p-2} ds}.
$$
We now use a simple change of variables to obtain   
$$
\xrm^{\eps}_{\eps}= \xrm(X_{\tau_{j}-},\delta_{j})=X_{\tau_{j}},
$$
in which $\xrm$ is defined in    \reff{eq: def xrm}, 
while
$$
\sup_{0\le t\le \eps }\E\left[|X_{\tau_{j}+t}- X_{\tau_{j}}|^{p}\right]\le C\eps^{\frac{p}{2}}.
$$
 Since $X$ and $\Se$ have the same dynamics on $[\tau_{j}+\eps,\tau_{ j+1})$, this shows that
\begin{align*}
\E\left[\sup_{[\tau_{j}+\eps,\tau_{j+1}) }|X_{t}- \Se_{t}|^{p}\right] &\le  C\E\left[ |X_{\tau_{j}+\eps}- \Se_{\tau_{j}+\eps}|^{p}\right]
\\
&\le  C\E\left[ |\xrm_{\eps}^{\eps}- \Se_{\tau_{j}+\eps}|^{p}+ |X_{\tau_{j}+\eps}- X_{\tau_{j}}|^{p}\right]\\
&\le C \Esp{|\Se_{\tau_{j}-}-X_{\tau_{j}-}|^{p}+\int_{0}^{\eps} |\Se_{\tau_{j}+s}- \xrm^{\eps}_{s}|^{p-2} ds + \eps^{\frac{p}{2}} } .
\end{align*}
For $p=2$, this provides 
\b*
 \Esp{\sup_{ [\tau_{j-1}+\eps,\tau_{j})} |X -X^{\eps} |^{p}+\sup_{0\le s\le \eps}\E[  |X_{\tau_{j}+s}-X^{\eps}_{\tau_{j}+\eps}|^{p}} \le C\eps^{\frac{p}{2}},
\e*
 by induction over $j$, and the case $p=4$ then follows from  the above.
For later use, note that the estimate
\be\label{eq: estimate Se - Xe}
\sup_{0\le t\le \eps }\E\left[|\Se_{\tau_{j}+t}- \xrm^{\eps}_{t}|^{4}\right]\le C  \eps^{2}
\ee
is a by-product of our analysis. 

b. The estimate on $V-\Ve$ is proved similarly. We introduce 
\b*
{\rm v}^{\eps}_{t}:=V_{\tau_{j}-}+ \int_{0}^{t}   \frac{\delta_{j}^{2}}{\eps^{2}} s f(\xrm^{\eps}_{s}) ds
+ Y_{\tau_{j}-}  \int_{0}^{t}   \frac{\delta_{j}}{\eps} f(\xrm^{\eps}_{s}) ds=V_{\tau_{j}-}+\int_{0}^{t} Y^{\eps}_{s}  \frac{\delta_{j}}{\eps} f(\xrm^{\eps}_{s}) ds,
\e*
and obtain a first estimate by using  \reff{eq: estimate Se - Xe}: 
\begin{align*}
\E\left[|V^{\eps}_{\tau_{j}+t}- {\rm v}^{\eps}_{t}|^{2}\right]&\le C \Esp{|V^{\eps}_{\tau_{j}-}-V_{\tau_{j}-}|^{2} +\eps + \left(\int_{0}^{\eps}\eps^{-1}Y^{\eps}_{\tau_{j}+s}\delta_{j}|X^{\eps}_{\tau_{j}+s}-\xrm^{\eps}_{s}| ds\right)^{2} }\\
&\le C \Esp{|V^{\eps}_{\tau_{j}-}-V_{\tau_{j}-}|^{2} +\eps },
\end{align*}
for $0\le t\le \eps$.
Then, we observe that 
\b*
{\rm v}^{\eps}_{\eps}=V_{\tau_{j}-}+  \J(X_{\tau_{j}-},\delta_{j}) +Y_{\tau_{j}-}  \I(X_{\tau_{j}-},\delta_{j}) =V_{\tau_{j}},
\e* 
while
$$
\sup_{0\le t\le \eps }\E\left[|V_{\tau_{j}+t}- V_{\tau_{j}}|^{2}\right]\le C\eps .
$$
By using the estimate on $X-X^{\eps}$ obtained in a., we then show that 
\b*
\E\left[\sup_{[\tau_{j}+\eps,\tau_{j+1}) }|V_{t}- V^{\eps}_{t}|^{2} \right] &\le&  C\E\left[ |V_{\tau_{j}+\eps}- V^{\eps}_{\tau_{j}+\eps}|^{2} +\eps \right],
\e*
and conclude  by using an induction over $j$. 

\noindent Step 2. We now consider the general case. We define
$$
\tau^{\eps}_{j+1}=(\eps+\tau^{\eps}_{j})\vee \tau_{j+1}\;,\; \delta^{\eps}_{j+1}=\int_{(\tau^{\eps}_{j},\tau^{\eps}_{j+1}]} \delta \nu(d\delta,dt)\;,\;j\ge 1,
$$
where $(\tau^{\eps}_{1},\delta^{\eps}_{1})=(\tau_{1},\delta_{1})$. On $E_{\eps}:=\{\min_{j\le k-1} (\tau_{j+1}-\tau_{j})\ge \eps\}$, $(\tau^{\eps}_{j},\delta^{\eps}_{j})_{j\ge 1}$ $=$  $(\tau_{j},\delta_{j})_{j\ge 1}$. Hence, it follows from Step 1. that
\b*
 \Esp{  |Z^{\eps}_{ T+\eps}-Z_{  T}|^{2}  }\le  C\eps +C\Esp{   |\tilde Z^{\eps}_{ T+\eps}|^{4}+|Z_{  T}|^{4}   }^{\frac12}\P[E_{\eps}^{c}]^{\frac12},
\e*
in which $\tilde Z^{\eps}$ stands for the dynamics associated to $(\tau^{\eps}_{j},\delta^{\eps}_{j})_{j\ge 1}$. It now follows from standard estimates that $(\tilde Z^{\eps}_{ T+\eps})_{0<\eps\le 1}$ and $Z_{T}$ are bounded in $\Lb^{4}$.
\ep
\\

We conclude this section with a  proposition collecting some important properties of the functions $\xrm$ and $\J$ which appear in Proposition \ref{prop : lim quand n to infty}. They will be used in the subsequent section.
\begin{Proposition}\label{prop: xrm and J} For all  $x,y,\iota \in \R$,
\begin{enumerate}[{\rm (i)}]
\item $\xrm(\xrm(x,\iota),-y-\iota)=\xrm(x,-y)$,
\item $f(x)\partial_{x}\xrm(x,y)=\partial_{y}\xrm(x,y)=f(\xrm(x,y))$,
\item   $\J(\xrm(\xrm(x,\iota),-y-\iota),y+\iota)-\J(\xrm(x,-y),y)=y\I(x,\iota)+\J(x,\iota)$,
\item $f(x)\partial_{x}\J(x,y)+\I(x,y)=\partial_{y}\J(x, y)=yf(\xrm(x,y))$.
\end{enumerate}
\end{Proposition}

\proof (i) is an immediate consequence of the Lipschitz continuity of the function $f$, which ensures uniqueness of the ODE defining $\xrm$ in \reff{eq: def xrm}. More generally, it has the flow property, which we shall use in the following arguments.  The assertion (ii) is an immediate consequence of the definition of $\xrm$: $\xrm(\xrm(x,\iota),y-\iota)=\xrm(x,y)$ for $\iota>0$ and $\partial_{y}\xrm(x,0)=f(x)$, so that differentiating at $\iota=0$ provides (ii). The identity in (iii) follows from direct computations. As for (iv), it suffices to write that
$\J(\xrm(x,\iota),y-\iota)=\int_{\iota}^{y} (t-\iota)f(\xrm(x,t))dt$ for $\iota>0$, and again to differentiate at $\iota=0$. \ep

\begin{Remark} It follows from Proposition \ref{prop: xrm and J} that our model allows round trips at (exactly) zero cost. Namely, if $x$ is the current stock price, $v$ the wealth, and $y$ the number of shares in the portfolio, then performing an immediate jump of size $\delta$ makes $(x,y,v)$ jump to $(\xrm(x,\delta)$, $y+\delta,$ $v+y\I(x,\delta)+\J(x,\delta))$. Passing immediately the opposite order, we come back to the position
$(\xrm(\xrm(x,\delta),-\delta)$, $y+\delta-\delta,$ $v+y\I(x,\delta)+\J(x,\delta)+(y+\delta)\I(\xrm(x,\delta),-\delta)+\J(\xrm(x,\delta),-\delta))$ $=$ $(x,y,v)$, 
by Proposition \ref{prop: xrm and J}(i)-(iii). This is a desirable property if one wants to have a chance to hedge options perfectly, or more generally to obtain a non-degenerated super-hedging price. 
\end{Remark}

%%%%%%%%%%%%%%%%%%%%%%%%%%%%%%%%%%%%%%%%%%%%%%%%%%%%%%%%%%%%%%%%%%%%%%%
%%%%%%%%%%%%%%%%%%%%%%%%%%%%%%%%%%%%%%%%%%%%%%%%%%%%%%%%%%%%%%%%%%%%%%%

\section{Super-hedging of a European claim}\label{sec: super hedging}

We now turn to the super-hedging problem. From now on, we define the admissible  strategies as the It\^{o} processes of the {form}
\be\label{eq: dyna Y adm}
Y=y+\int_{0}^{\cdot} b_{s} ds +\int_{0}^{\cdot} a_{s} dW_{s}  +\int_{0}^{\cdot}\int\delta\nu(d\delta,ds)
\ee
in which $y\in \R$,  $(a,b,\nu)\in \Ac\x \Uc$  and $Y$  is essentially bounded. If $|Y|\le k$ and $(a,b,\nu)\in \Ac_{k}\x \Uc_{k}$, then we say that $(a,b,\nu)\in \Gamma_{k}$, $k\ge 1$, and we let $$\Gamma:=\cup_{k\ge 1} \Gamma_{k}.$$
We will comment in Remark \ref{rem: controle borne} below the reason why we restrict to bounded controls.

 Given $(t,z)\in  \D:=[0,T]\x \X\x \R\x \R$, we   define  $$Z^{t,z,\gamma}:=(X^{t,z,\gamma},Y^{t,z,\gamma},V^{t,z,\gamma})$$ as the solution of \reff{eq: S lim conti avec jump}-\reff{eq: dyna Y adm}-\reff{eq: V lim conti avec jump} on $[t,T]$ associated to $\gamma \in \Gamma$ and with initial condition $ Z^{t,z,\gamma}_{t-}$ $=$ $z$.

\subsection{Super-hedging price}

 A European contingent claim is defined by its payoff function, a measurable map  $x\in \X\mapsto (g_{0},g_{1})(x)\in \R^{2}$. The first component is the cash-settlement part, i.e.~the amount of cash paid at maturity, while $g_{1}$ is the delivery part, i.e. the number of units of stocks to be delivered.

An admissible strategy $\gamma\in \Gamma$ allows to super-hedge the claim associated to the payoff $g$, starting from the initial conditions $z$ at time $t$ if
$$
Z^{t,z,\gamma}_{T} \in \G
$$
where
\be\label{eq: def G}
\G:=\{(x,y,v)\in  \X\x \R\x \R~:~v-yx\ge g_{0}(x) \mbox{ and } y= g_{1}(x)\}.
\ee
Recall that $V$ stands for the frictionless liquidation value of the portfolio, it is the sum of the cash component and the value $YX$ of the stocks held without taking the liquidation impact  into account.

We set
$$
\Gc_{k}(t,z):=\{\gamma \in \Gamma_{k}~:~ Z^{t,z,\gamma}_{T}\in \G\}\;,\;\Gc(t,z):=\cup_{k\ge 1} \Gc_{k}(t,z),
$$
and  define the super-hedging price as
\b*
&w(t,x):=\inf_{k\ge 1} w_{k}(t,x)\;\mbox{ where }\; w_{k}(t,x):= \inf\{v:  \Gc_{k}(t,x,0,v)\ne \emptyset\}.\;&\;
\e*

For later use, let us make precise what are the $T$-values of these functions. 
\begin{Proposition}%\label{rem: w(T,cdot)} 
Define 
\begin{align*}
G_{k}(x)&:=\inf\{y\xrm(x,y)+ g_{0}(\xrm(x,y))-\J(x,y): |y|\le k \mbox{ s.t. } y= g_{1}(\xrm(x,y)) \}, \;x\in \R, 
\end{align*}
and $G:= \inf_{k\ge 1} G_{k}$. Then, 
\be\label{eq: w(T,x)}
w_{k}(T,\cdot)=G_{k} &\mbox{ and }&w(T,\cdot)=G.
\ee
\end{Proposition}
\proof Set $z=(x,0,v)$ and fix $\gamma=(a,b,\nu)\in \Gamma$. By \reff{eq: S lim conti avec jump}-\reff{eq: V lim conti avec jump}, we have 
\b*
Z^{T,z,\gamma}_{T}=(\xrm(x,y),y,v+\J(x,y))\;\mbox{ with } y:=\int \delta \nu(d\delta,\{T\}).
\e* 
In view of  \reff{eq: def G},  $Z^{T,z,\gamma}_{T}\in \G$ is then equivalent to 
$$
v+\J(x,y)-y\xrm(x,y)\ge g_{0}(\xrm(x,y)) \;\mbox{ and }\; y=g_{1}(\xrm(x,y)).  
$$
By definition of $w$ (resp. $w_{k}$), we have to compute the minimal $v$ for which this holds for some $y\in \R$ (resp. $|y|\le k$).
 \ep 

\begin{Remark}\label{rem: controle borne} Let us conclude this section with a comment on our choice of the set of  bounded controls $\Gamma$.

a. First, this ensures that the dynamics of $X, Y$ and $V$ are well-defined. This could obviously be relaxed by imposing $\Lb^{2}_{\lambda}$ bounds. However, note that the bound should anyway be uniform. This is crucial to ensure that the dynamic programming principle stated in
Section \ref{sec : dpp} is valid, as it uses measurable selection arguments: $\omega\mapsto \vartheta[\omega]\in \Lb^{\lambda}_{2}$  does not imply $\Esp{\|\vartheta[\cdot]\|_{\Lb_{2}^{\lambda}}}<\infty$. See Remark \ref{rem : dpp wk} below for a related discussion.

b. In the proof of Theorem \ref{thm : pde}, we will need to perform a change of measure associated to a martingale of the form $dM=-M\chi^{a}dW$ in which $\chi^{a}$ may explode  at a speed $a^{2}$ if    $a$ is not bounded. See Step 1.~of the proof of Theorem \ref{thm : pde}. In order to ensure that this local martingale is well-defined, and is actually a martingale, one should impose very strong integrability conditions on $a$.

In order to simplify the presentation, we therefore stick to bounded controls. Many other choices are possible. Note however that, in the case $f\equiv 0$, a large class of options  leads to hedging strategies in our set $\Gamma$, up to  a slight payoff smoothing to avoid the explosion of the delta or the gamma at maturity. This implies that, although the perfect hedging strategy may not belong to $\Gamma$, at least it is a limit of elements of $\Gamma$ and the super-hedging prices coincide.
\end{Remark}
%%%%%%%%%%%%%%%%%%%%%%%%%%%%%%%%%%%%%%%%%
\subsection{Dynamic programming}\label{sec : dpp}

Our control problem is a stochastic target problem as studied in \cite{soner2002dynamic}. The aim of this section is to show that it satisfies a version of their geometric dynamic programming principle.

However, the value function $w$ is not amenable to dynamic programming per se. The reason is that it assumes a zero initial stock holding at time $t$, while the position $Y_{\theta}$ will (in general) not be zero at a later time $\theta$. It is therefore a priori not possible to compare the later wealth process $V_{\theta}$ with the corresponding super-hedging price $w(\theta,X_{\theta})$.

Still, a version of the geometric dynamic programming principle can be obtained if we introduce the process
\be\label{eq: def hat X}
\hat X^{t,z,\gamma}:=\xrm(X^{t,z,\gamma},-Y^{t,z,\gamma})
\ee
which represents the value of the stock immediately after liquidating the stock position.

We refer to Remark \ref{rem : dpp wk} below for the reason why part (ii) of the following dynamic programming principle is stated in terms of $(w_{k})_{k\ge 1}$ instead of $w$. 

\begin{Proposition}[GDP]\label{prop: DPP}  Fix $(t,x,v)\in [0,T]\x\X\x\R $.
\begin{enumerate}[\rm(i)]
\item If $v>w(t,x)$ then there exists $\gamma\in \Gamma$ and $y\in \R$ such that
 \b*
 V^{t,z,\gamma}_{\theta}&\ge& w(\theta,\hat X^{t,z,\gamma}_{\theta}) + \J( \hat X^{t,z,\gamma}_{\theta} , Y^{t,z,\gamma}_{\theta}),
 \e*
 for all stopping time $\theta\ge t$,  where $z:=(\xrm(x,y),y,v+\J(x,y))$.
 \item Fix $k\ge 1$.    If $v<w_{2k+2}(t,x)$ then we can not find $\gamma\in \Gamma_{k}$, $y\in [-k,k]$ and a stopping time $\theta\ge t$ such that
 \b*
 &V^{t,z,\gamma}_{\theta}> w_{k}(\theta,\hat X^{t,z,\gamma}_{\theta})  + \J( \hat X^{t,z,\gamma}_{\theta} , Y^{t,z,\gamma}_{\theta})&
 \e*
 with   $z:=(\xrm(x,y),y,v+\J(x,y))$.
\end{enumerate}

\end{Proposition}

\proof Step 1. In order to transform our stochastic target problem into a time consistent one, we introduce the auxiliary value function corresponding to an initial holding $y$ in stocks:
\b*
&\hat w(t,x,y):=\inf_{k\ge 1} \hat w_{k}(t,x,y)\;\mbox{ where }\; \hat w_{k}(t,x,y):= \inf\{v:  \Gc_{k}(t,x,y,v)\ne \emptyset\}.\;&\;
\e*
Note that $w_{k+1}(t,x)\le \inf\{v: \exists\; y\in [-k,k]$ s.t.~$\Gc_{k}(t,\xrm(x,y),y,v+\J(x,y))\ne \emptyset\}$. This follows from \reff{eq: S lim conti avec jump}-\reff{eq: V lim conti avec jump}. Since $\xrm(\xrm(x,-y),y)=x$, see  Proposition \ref{prop: xrm and J}, this implies that
\be\label{eq: hat w ge w}
\hat w_{k}(t,x,y)\ge w_{k+1}(t,\xrm(x,-y))+\J(\xrm(x,-y),y),
\ee
for $|y|\le k$.
Similarly,  since $\J(x,-y)+y\I(x,-y)=-\J(\xrm(x,-y),y)$ by Proposition \ref{prop: xrm and J}, we have
\be\label{eq: hat w le w}
\hat w_{k+1}(t,x,y)\le w_{k}(t,\xrm(x,-y))+\J(\xrm(x,-y),y).
\ee

\noindent {Step 2.} {a.} Assume that $v>w(t,x)$. The  definition of $w$ implies that  we can find $y\in \R$ and $\gamma\in \Gc(t,z)$ where $z:=(\xrm(x,y),y,v+\J(x,y))$. By the arguments of \cite[Step 1 proof of Theorem 3.1]{soner2002dynamic},
 $ V^{t,z,\gamma}_{\theta}$ $\ge$ $ \w(\theta,X^{t,z,\gamma}_{\theta},Y^{t,z,\gamma}_{\theta})$, for all stopping time $\theta\ge t$.
Then, \reff{eq: hat w ge w} applied  for $k\to \infty$ provides (i).

{b.}     Assume now that  we can  find   $\gamma\in \Gamma_{k}$, $y\in [-k,k]$ and a stopping time $\theta\ge t$ such that  $V^{t,z,\gamma}_{\theta}$ $>$ $ (w_{k}+\J)(\theta,\hat X^{t,z,\gamma}_{\theta},Y^{t,z,\gamma}_{\theta})$, where $z:=(\xrm(x,y),y,v+\J(x,y))$. By \reff{eq: def hat X}-\reff{eq: hat w le w}, $V^{t,z,\gamma}_{\theta}$ $>$ $ \w_{k+1}(\theta,X^{t,z,\gamma}_{\theta},Y^{t,z,\gamma}_{\theta})$, and it follows from \cite[Step 2 proof of Theorem 3.1]{soner2002dynamic} and Corollary \ref{cor : measurability} that $v+\J(x,y)\ge \w_{2k+1}(t,\xrm(x,y),y)$. We conclude that (ii) holds by appealing to \reff{eq: hat w ge w} and the identities  $\xrm(\xrm(x,y),-y)=x$ and $\J(\xrm(\xrm(x,y),$ $-y),$ $y)$ $=$ $\J(x,y)$, see Proposition \ref{prop: xrm and J}.
\ep\\

We conclude this section with purely technical considerations that justify the form of the above dynamic programming principle. They are of no use for the later developments but may help to clarify our approach. 
\begin{Remark}\label{rem : dpp wk} Part {\rm (ii)} of Proposition \ref{prop: DPP} can not be stated in terms of $w$. The reason is that measurable selection technics can not be used with the set $\Gamma$. Indeed, if $\omega\mapsto \gamma[\omega]\in \Gamma$, then the corresponding bounds depend on $\omega$ and are not uniform: a measurable family of controls $\{\gamma[\omega],\omega\in \Omega\}$ does not permit to construct an element in $\Gamma$. Part {\rm (i)} of Proposition \ref{prop: DPP} only requires to use a conditioning argument, which can be done within $\Gamma$.
\end{Remark}

\begin{Remark}\label{rem : pb sing pour hat w} A version of the geometric dynamic programming principle also holds for $(\hat w_{k})_{k\ge 1}$, this is a by-product of the above proof. It is therefore tempting to try to derive a pde for the function $\hat w$. However, the fact that the control $b$ appears linearly in the dynamics of $(X,Y,V)$ makes this problem highly singular, and ``standard approaches'' do not seem to work.  We shall see in Lemma \ref{lem : ito V-w}   that this singularity disappears in the parameterization  $\xrm(X,-Y)$ used in Proposition \ref{prop: DPP}.  Moreover, hedging implies a control on the diffusion part of the dynamics which translates into a strong relation between $Y$ and the space gradient $D\hat w(\cdot,X,Y)$. This would lead to a pde set on a curve on the coordinates $(t,x,y)$ depending on $D\hat w$ (the solution of the pde). 
\end{Remark}
%%%%%%%%%%%%%%%%%%%%%%%%%%%%%%%%%%%%%%%%
\subsection{Pricing equation}

In order to understand what is the partial differential equation that $w$ should solve, let us state the following key lemma. Although the control $b$ appears linearly in the dynamics of $(X,Y,V)$, the following shows that the singularity this may create does indeed not appear when applying It\^{o}'s Lemma to $V-(\vp+\J)(\cdot,\hat X,Y)$, recall \reff{eq: def hat X}, it is absorbed by the functions $\xrm$ and $\J$ (compare with Remark \ref{rem : pb sing pour hat w}).   The proof of this Lemma  is postponed to Section \ref{sec: proofs}.  

\begin{Lemma}\label{lem : ito V-w} Fix $(t,x,y,v)\in \D$, $z:=(x,y,v)$,    $\gamma=(a,b,\nu) \in \Gamma$. Then,
\b*
\hat X^{t,z,\gamma}&=&\xrm(x,-y)\\
&+& \int_{t}^{\cdot}[ \hat \mu(\hat X^{t,z,\gamma}_{s},Y^{t,z,\gamma}_{s})+(\partial_{x} \xrm \mu-\frac12\partial_{x} \xrm  a^{2}_{s}ff')(X^{t,z,\gamma}_{s},-Y^{t,z,\gamma}_{s})]ds\\
&+& \int_{t}^{\cdot}\hat \sigma(\hat X^{t,z,\gamma}_{s},Y^{t,z,\gamma}_{s})dW_{s}.
\e*
Given $\vp\in C^{\infty}_{b}$, set $\Ec^{t,z,\gamma}:= V^{t,z,\gamma}-(\vp+\J)(\cdot,\hat X^{t,z,\gamma},Y^{t,z,\gamma})$. Then,
\b*
\Ec^{t,z,\gamma}-\Ec^{t,z,\gamma}_{t}&=&  \int_{t}^{\cdot }
 [Y^{t,z,\gamma}_{s}-\check Y^{t,z,\gamma}_{s} ](\mu- f'f a_{s}^{2}/2) (X^{t,z,\gamma}_{s})ds\\
&&+\;\int_{t}^{\cdot}[Y^{t,z,\gamma}_{s}-\check Y^{t,z,\gamma}_{s}]\sigma(X^{t,z,\gamma}_{s})dW_{s}  \\
&&+ \;\int_{t}^{\cdot}\hat F\vp(s,\hat X^{t,z,\gamma}_{s},Y^{t,z,\gamma}_{s})ds
\e*
in which
\b*
\check Y^{t,z,\gamma}&:=& Y^{t,z,\gamma} + \frac{\hat X^{t,z,\gamma}- X^{t,z,\gamma}}{f(  X^{t,z,\gamma})} +  \partial_{x}\vp(\cdot,\hat X^{t,z,\gamma})\frac{f(\hat X^{t,z,\gamma})}{f(X^{t,z,\gamma})}
\e* 
and
\b*
\hat F\vp&:=&-\partial_{t}\vp - \hat \mu  \partial_{x}[\vp +\J]-\frac12 \hat \sigma^{2}  \partial^{2}_{xx}[\vp+\J],
\e*
where for $(x',y')\in \X\x \R$
$$
\hat \mu(x',y'):= \frac12 [\partial^{2}_{xx}\xrm \sigma^{2} ](\xrm(x',y'),-y')
\;\mbox{ and }\, \hat \sigma(x',y'):=(\sigma\partial_{x} \xrm)(\xrm(x',y'),-y').
$$
\end{Lemma}

Let us now appeal to Proposition \ref{prop: DPP} and apply Lemma \ref{lem : ito V-w}  to $\vp=w$,  assuming that $w$ is smooth and that  Proposition   \ref{prop: DPP}(i) is valid even if we start from $v=w(t,x)$, i.e.~assuming that the $\inf$ in the definition of $w$ is a $\min$. With the notations of the above lemma, 
 Proposition   \ref{prop: DPP}(i) formally applied to $\theta=t+$ leads to
\b*
0&\le& d\Ec^{t,z,\gamma}_{t}\\
&=&(y-\hat y)\left\{[   \mu-  ff'a^{2}_{t}/2)(\xrm(x,y))]dt+ \sigma(\xrm(x,y)) dW_{t}\right\}\\
&+& \hat Fw(t,\hat x,y)dt
\e*
in which
\b*
\hat y= y +  \frac{\hat x-\xrm(  x,y)}{f(\xrm(  x,y))} +  \partial_{x}w(t,\hat x)\frac{f(\hat x)}{f(\xrm( x,y))}
&\mbox{ and }&
\hat x= \xrm(\xrm(x,y),-y)=x.
\e*
Remaining at a formal level, this inequality cannot hold unless $y=\hat y$, because $\sigma\ne 0$, and
$$
\hat Fw(t,x,\hat y)=\hat Fw(t,\hat x,y)\ge  0.
$$
This means that $w$ should be a super-solution of
\be\label{eq: PDE}
F\vp(t,x)&:=&\hat F\vp(t,x,\hat y[\vp](t,x))=0
\ee
where, for a smooth function $\vp$,
$$
\hat y[\vp](t,x):=\xrm^{-1}(x,x+f(x)\partial_{x}\vp(t,x))
$$
and $\xrm^{-1}$ denotes the inverse of $\xrm(x,\cdot)$.

From (ii) of Proposition \ref{prop: DPP}, we can actually (formally) deduce that the above inequality should be an equality, and therefore that
$w$ should solve  \reff{eq: PDE}.

In order to give a sense to the above, we assume that
\b*
\left\{
\begin{array}{c}
\xrm(x,\cdot) \mbox{ is invertible for all $x\in \X$}
\\
(x,z)\in \X\x \R\mapsto \xrm^{-1}(x,z) \;\mbox{ is $C^{2}$.  }
\end{array}
\right. &&{\bf \rm (H2)}
\e*

In view of \reff{eq: w(T,x)},  we therefore expect $w$ to be a solution of
\be\label{eq: PDE up to T}
F\vp \1_{[0,T[}+(\vp-G)\1_{\{T\}}=0\;\mbox{ on } [0,T]\x \X.
\ee
Since $w$ may not be smooth and (ii) of Proposition \ref{prop: DPP} is stated in terms of $w_{k}$ instead of $w$, we need to consider the notion of viscosity solutions and the relaxed semi-limits of $(w_{k})_{k\ge 1}$. We therefore define
$$
w_{*}(t,x):=\liminf_{(t',x',k)\to (t,x,\infty)} w_{k}(t',x') \mbox{ and }  w^{*}(t,x):=\limsup_{(t',x',k)\to (t,x,\infty)} w_{k}(t',x'),
$$
in which the limits are taken over $t'<T$, as usual. Note that $w_{*}$ actually coincides with the lower-semicontinuous enveloppe of $w$, this comes from the fact that   $w=\inf_{k\ge 1} w_{k}=\lim_{k\to \infty} \downarrow w_{k}$, by construction.
\vs2

 We are now in position to state the main result of this section.  In the following, we  assume that
\b*
\left\{\begin{array}{c}G \;\mbox{ is continuous and }\;G_{k}\downarrow G\; \mbox{ uniformly on compact sets.}\\
\mbox{$w_{*}$ and $w^{*}$ are finite on $[0,T]\x \X$.}
\end{array}
\right. &&{\bf \rm (H3)}
\e*
 The first part of {\rm (H3)} will be used to obtain the boundary condition. The second part is natural since otherwise our problem would be ill-posed.

 \begin{Theorem}[Pricing equation]\label{thm : pde} The functions $w_{*}$ and $w^{*}$ are respectively a viscosity super- and a subsolution of \reff{eq: PDE up to T}.  If they are bounded and $\inf f>0$, then  $w=w_{*}=w^{*}$  and $w$ is the unique bounded viscosity solution of \reff{eq: PDE up to T}. If in addition  $G$ is bounded and $C^{2}$ with $G,G',G^{''}$ H\"older continuous, then $w \in C^{1,2}([0,T)\x \R)\cap C^{0}([0,T]\x \R)$.  
 \end{Theorem}
 
The proof is reported in Section \ref{sec: proofs}. 
 Let us now discuss the verification counterpart. 
 \begin{Remark}[Verification]\label{rem : verification} Assume that $\vp$ is a smooth solution of \reff{eq: PDE up to T} and that we can find $(a,b)\in \Ac$ such that the following system holds on $[t,T)$:
 \b*
  X&=&x+\I(x,\hat y[\vp](t,x))+\int_{t}^{\cdot} \sigma(X_{s}) dW_{s} + \int_{0}^{\cdot} f(X_{s}) d Y^{c}_{s}\\
&&+\;\int_{0}^{\cdot} (\mu(X_{s})+a_{s}(\sigma f')(X_{s}) )ds+\I(X_{T-},-Y_{T-})\1_{\{T\}}
\\
  Y&=&\hat y[\vp](t,x)+\int_{t}^{\cdot} b_{s} ds + \int_{t}^{\cdot} a_{s} dW_{s}-Y_{T-}\1_{\{T\}}\\
  &=&\xrm^{-1}(\hat X,\hat X+(f\partial_{x}\vp)(\cdot,\hat X))-Y_{T-}\1_{\{T\}}\\
  \hat X&:=& \xrm(X,-Y)\\
  V&=&\vp(t,x)+\J(x,\hat y[\vp](t,x) )+ \int_{t}^{\cdot} Y_{s}dX^{c}_{s}+\frac12 \int_{0}^{\cdot}  a_{s}^{2} f(X_{s}) ds\\
  &&+\; (Y_{T-}\I(X_{T-},-Y_{T-})+\J(X_{T-},-Y_{T-}))\1_{\{T\}}.
\e*
 a. Note that $\hat X_{t}=\xrm(X_{t},-Y_{t})=\xrm(\xrm(x,\hat y[\vp](t,x)),-\hat y[\vp](t,x))=x$, recall Proposition  \ref{prop: xrm and J}(i), so that 
$Y_{t}=\hat y[\vp](t,x)$ $=$ $\xrm^{-1}(\hat X_{t},\hat X_{t}+(f\partial_{x}\vp)(t,\hat X_{t}))$. We therefore need to find $(a,b)$ such that 
$X=\xrm(\hat X,Y)=\hat X+(f\partial_{x}\vp)(\cdot,\hat X)$. This amounts to solving:
\begin{align*}
\sigma(X)+f(X)a &= \hat \sigma(\hat X,Y) \partial_{x} \psi(\cdot,\hat X)\\
f(X)b+(\mu+a\sigma f')(X)&=(\hat \mu(\hat X,Y)+ (\partial_{x} \xrm \mu-\frac12\partial_{x} \xrm  a^{2}_{s}ff')(X,-Y))\partial_{x}\psi(\cdot,X)\\
&+ \frac12 \hat \sigma^{2}(\hat X,Y) \partial^{2}_{xx} \psi(\cdot,\hat X)
\end{align*}
where 
$\psi(t,x):=x+(f\partial_{x}\vp)(t,x)$. Since $f>0$, this system has a solution. Under additional smoothness and boundedness assumption, $(a,b)\in \Ac$.

b. Let $\check Y$ be as in  Lemma \ref{lem : ito V-w} for the above dynamics. 
Since   $X=\xrm(\hat X,Y)=\hat X+(f\partial_{x}\vp)(\cdot,\hat X)$ on $[t,T)$ by construction, we have $\check Y=Y$ on $[t,T)$.
Then, it follows from Lemma \ref{lem : ito V-w} and \reff{eq: PDE}-\reff{eq: PDE up to T} that
$$
V_{T-}=\vp(T,\hat X_{T-})+\J(\hat X_{T-},Y_{T-})=G(\hat X_{T-})+\J(\hat X_{T-},Y_{T-}).
$$
Since $X_{T}=\hat X_{T-}$ and $Y_{T-}\I(X_{T-},-Y_{T-})+\J(X_{T-},-Y_{T-})+\J(\hat X_{T-},Y_{T-})=0$, see Proposition \ref{prop: xrm and J}, this implies that $V_{T}$ $=$ $G(X_{T})$. Hence, the hedging strategy consists in taking an initial position is stocks equal to $Y_{t}=\hat y[\vp](t,x)$ and then to use the control $(a,b)$ up to $T$. A final immediate trade  is performed at $T$. In particular,  the number of stocks $Y$ is continuous on $(t,T)$.

%c. We refer to \cite{ladyzhenska1988linear} for conditions under which a smooth solution of \reff{eq: PDE up to T} exists and for estimates on its derivatives. 
 \end{Remark}

%%%%%%%%%%%%%%%%%%%%%%%%%%%%%
\subsection{An example: the fixed impact case}\label{sec: example}

In this section, we consider the simple case of a constant impact function $f$: $f(x)=\lambda>0$ for all $x\in \R$. This is certainly a too simple model, but this allows us to highlight the structure of our result as the pde simplifies in this case. Indeed, for  
\b*
\xrm(x,y)=x+y \lambda&\mbox{ and }& \J(x,y)=\frac12 y^{2}\lambda, 
\e*
we have 
$$
\hat \mu(x,y)=0
\;\mbox{ , }\, \hat \sigma(x,y):=\sigma(x+y\lambda) \;\mbox{ , }\, \hat y[\vp]:=\partial_{x}\vp.
$$
The pricing equation is given by a local volatility model in which the volatility depends on the hedging price itself, and therefore on the claim $(g_{0},g_{1})$ to be hedged:
$$
0=-\partial_{t}\vp(t,x) -  \frac12  \sigma^{2}(x+\partial_{x}\vp \lambda)  \partial^{2}_{xx}\vp(t,x).
$$
As for the process $Y$ in the verification argument of Remark \ref{rem : verification}, it is given by 
$$
Y=\partial_{x}\vp(\cdot,\hat X)=\partial_{x}\vp(\cdot,X-\lambda Y). 
$$ 
This shows that the hedging strategy (if it is well-defined) consists in following the usual $\Delta$-hedging strategy but for a $\Delta=\partial_{x}\vp$ computed at the value of the stock $\hat X$ which would be obtained if the position in stocks was liquidated. 
 \\

Note that we obtain the usual heat equation when $\sigma$ is constant. This is expected, showing the limitation of the fixed impact model. To explain this, let us consider the simpler case $g_{1}=0$ and use the notations of   Remark \ref{rem : verification}. We also set $\mu=0$ for ease of notations. Since $\sigma$ is constant, the strategy $Y$ does not affect the coefficients in the dynamics of $X$, it just produces a shift $\lambda dY$ each time we buy or sell. Since $Y_{T}=0$, and $Y_{t-}=0$, the total impact is null: $X_{T}=X_{t-}+\sigma (W_{T}-W_{t})$. 
As for the wealth process, we have 
\begin{align*}
V_{T}&=\vp(t,x)+\frac12 Y^{2}_{t}\lambda+\int_{t}^{T} Y_{s}dX^{c}_{s} +\frac12\int_{t}^{T} a^{2}_{s}\lambda ds - Y_{T-}^{2}\lambda+ \frac12 Y_{T-}^{2}\lambda
\\
&=\vp(t,x)+\int_{t}^{T} Y_{s}\sigma dW_{s}+\frac12 \lambda (Y^{2}_{t}-Y^{2}_{T-})+  \int_{t}^{T}\lambda Y_{s}dY^{c}_{s}+\frac12\int_{t}^{T} a^{2}_{s}\lambda ds\\
&=\vp(t,x)+\int_{t}^{T} Y_{s}\sigma dW_{s}.
\end{align*}
Otherwise stated, the liquidation costs are cancelled: when buying, the trader pays a costs but moves the price up, when  selling back, he pays a cost again but sell at a higher price. If there is no effect on the underlying dynamics of $X$ and $f$ is constant, this perfectly cancels. 

However, the hedging strategy is still affected: $Y=\partial_{x}\vp(\cdot,X-\lambda Y)$.

%%%%%%%%%%%%%%%%%%%%%
\subsection{Proof of   the pde characterization}\label{sec: proofs}

%%%%%%%%%%%%%%%%%%%%
\subsubsection{The key lemma}

We first provide the proof of our key result. 
\vs2

\noindent{\bf Proof of Lemma \ref{lem : ito V-w}.} To alleviate the notations, we omit the super-scripts. \\
 a. We   first observe from  Proposition \ref{prop: xrm and J}(i)   that $ \xrm(X,-Y)$ has continuous paths, while     Proposition \ref{prop: xrm and J}(ii) implies  that
$f \partial_{x}\xrm -\partial_{y}\xrm =0$ (and therefore $f' \partial_{x}\xrm +f \partial^{2}_{xx}\xrm -\partial^{2}_{xy}\xrm=0$). Using It\^{o}'s Lemma, this   leads  to
\b*
d\xrm(X_{s},-Y_{s})&=& (\mu-\frac12 a^{2}_{s}ff')(X_{s})\partial_{x} \xrm(X_{s},-Y_{s})ds+\sigma(X_{s})\partial_{x} \xrm(X_{s},-Y_{s})dW_{s}\\
&&+\frac12\left[ \sigma^{2}\partial^{2}_{xx}\xrm -  a^{2}_{s}f\partial^{2}_{xy}\xrm+a^{2}_{s}\partial^{2}_{yy}\xrm \right](X_{s},-Y_{s})ds.
\e*
We now use the identity $f\partial^{2}_{xy}\xrm-\partial^{2}_{yy}\xrm=0$, which also follows from    Proposition \ref{prop: xrm and J}(ii), to simplify the above expression into
\b*
d\xrm(X_{s},-Y_{s})&=&[ \partial_{x} \xrm(\mu -\frac12  a^{2}_{s}ff' )+\frac12 \partial^{2}_{xx}\xrm \sigma^{2} ](X_{s},-Y_{s})ds\\
&&+\; (\sigma\partial_{x} \xrm)(X_{s},-Y_{s})dW_{s}.
\e*
{b.} Similarly, it follows  from  Proposition \ref{prop: xrm and J}(iii) that $V-\J (\hat X,Y)$ has continuous paths, and so does $\Ec$ by a.
 Before to apply It\^{o}'s lemma to derive the dynamics of $\Ec$, let us observe that $\partial_{y}\J(\xrm(x,-y),y)=yf(\xrm(\xrm(x,-y),y))$ $=$ $yf(x)$ and that $\partial^{2}_{yy}\J(\xrm(x,-y),y)=y(ff')(x)+f(x)$. Also note that $\hat \sigma(\xrm(x,-y),y)=\sigma(x)\partial_{x}\xrm(x,-y)$.  Then, using the dynamics of $\hat X$ derived above, we obtain
\begin{align*}
d\Ec_{s} =&
 (Y_{s}-\check Y_{s})\sigma(X_{s}) dW_{s } + (Y_{s}-\check Y_{s})[\mu-\frac12 a^{2}_{s}(ff') ](X_{s})ds
+\hat F\vp(s,\hat X_{s},Y_{s})ds\\
&+\;a_{s}\sigma(X_{s}) [Y_{s}f'(X_{s})-    \partial_{x}\xrm(X_{s},-{Y_{s}})\partial^{2}_{xy}\J(\hat X_{s},Y_{s})]ds,
\end{align*}
where 
$$
\check Y:=\partial_{x}(\vp+\J)(\cdot,\hat X,Y)\partial_{x}\xrm(X,-Y).
$$
By  Proposition \ref{prop: xrm and J}(ii)(iv), $
f(x)\partial^{2}_{xy}\J(x, y)=\partial_{y}[yf(\xrm(x,y))-\I(x,y)]=y (f'f)(\xrm(x,y))
$.
Since $\partial_{x}\xrm(x,-y)=f(\xrm(x,-y))/f(x)$, see  Proposition \ref{prop: xrm and J}(ii), it follows that
$$
 \partial_{x}\xrm(X,-{Y})\partial^{2}_{xy}\J(\xrm(X,-Y),Y) =Yf'(X),
$$
which implies   
$$
d\Ec_{s} =
 (Y_{s}-\check Y_{s})\sigma(X_{s}) dW_{s } + (Y_{s}-\check Y_{s})[\mu-\frac12 a^{2}_{s}(ff') ](X_{s})ds
+\hat F\vp(s,\hat X_{s},Y_{s})ds.
$$
We now deduce from Proposition \ref{prop: xrm and J} that 
\b*
\partial_{x} \J(\hat X,Y)&=&\frac{-\I(\hat X,Y)+Yf(\xrm(\hat X,Y))}{f(\hat X)}=\frac{\hat X-X +Y f(X)}{f(\hat X)}\\
\partial_{x}\xrm(X,-Y)&=&f(\hat X)/f(X),
\e*
so that 
$$
\check Y=\partial_{x}\vp(\cdot,\hat X)\frac{f(\hat X)}{f(X)} +\frac{\hat X-X}{f( X)}+Y.
$$
\ep
%
%\begin{Remark} \red{We have seen in the course of the proof of Lemma \ref{lem : ito V-w} that  $f \partial_{x}\xrm -\partial_{y}\xrm =0$ (and therefore $f' \partial_{x}\xrm +f \partial^{2}_{xx}\xrm -\partial^{2}_{xy}\xrm=0$). Since $\partial_{y}\xrm=f\circ \xrm$, this implies that
%\b*
% \partial_{x}\xrm=(f\circ \xrm)/f &\mbox{ and }& \partial^{2}_{xx}\xrm =( f\circ \xrm) (f'\circ \xrm - f')/f^{2}.
%\e*
%Since $\xrm(x,\hat y[\vp](t,x)) = x+f(x)\partial_{x}\vp(t,x) $, this implies that
%\b*
%\hat \mu(x,\hat y[\vp](t,x))&=& \frac12 [\partial^{2}_{xx}\xrm \sigma^{2} ](\xrm(x,\hat y),-\hat y) \\
%&=&\frac12 \frac{f(x)\left[f'(x)-f'(x+f(x)\partial_{x}\vp(t,x))\right]\sigma^{2}(x+f(x)\partial_{x}\vp(t,x) )}{f(x+f(x)\partial_{x}\vp(t,x) )^{2}} \\
%\hat \sigma(x,\hat y[\vp](t,x))&=&(\partial_{x} \xrm \sigma)(\xrm(x,\hat y),-\hat y)=\frac{\sigma\left(x+f(x)\partial_{x}\vp(t,x) \right)f(x)}{f(x+f(x)\partial_{x}\vp(t,x) )}
%\e*}
%\end{Remark}

%%%%%%%%%%%%%%%%%%%%%%%%
\subsubsection{Super- and subsolution properties}

We now prove the super- and subsolution properties of Theorem \ref{thm : pde}.
\vs2

\noindent {\bf Supersolution property.} We first prove the supersolution property. It follows from similar arguments as in \cite{bouchard2009stochastic}. Let $\vp$ be a $C^{\infty}_{b}$ function, and  $(t_{o},x_{o})\in [0,T]\x \X$ be a strict  (local) minimum point of $w_{*}-\vp$ such that $(w_{*}-\vp)(t_{o},x_{o})=0$.

a. We first assume that $t_{o}<T$ and $F\vp(t_{o},x_{o})<0$, and work towards a contradiction.  In view of \reff{eq: PDE},
\b*
\hat F\vp(t,x,y)<0\;\mbox{ if $(t,x)\in B$ and $|y-\hat y[\vp](t,x)|\le \eps$},
\e*
for some open ball $B\subset [0,T[\x \X$ which contains $(t_{o},x_{o})$, and some $\eps>0$.
Since $\xrm^{-1}$ is continuous, this implies that  
\begin{equation}\label{eq: sur sol eq contra t<T}
\hat F\vp(t,x,y)<0\;\mbox{ if $(t,x)\in B$ and  $|x+\partial_{x}\vp(t,x) f(x) -\xrm(x,y)|\le \eps f(\xrm(x,y))$},
\end{equation}
after possibly changing $B$ and $\eps$. 
 Let $(t_{n},x_{n})_{n}$ be a sequence in $B$ that converges to $(t_{o},x_{o})$ and such that $w(t_{n},x_{n})\to w_{*}(t_{o},x_{o})$ (recall that $w_{*}$ co\"incides with the lower-semicontinuous enveloppe of $w$). Set $v_{n}:=w(t_{n},x_{n})+n^{-1}$. It follows from Proposition \ref{prop: DPP}(i) that we can find  $(a^{n},b^{n},\nu^{n})=\gamma_{n}\in  \Gamma$ and $y_{n}\in \R$ such that
 \be\label{eq: proof sur sol Y ge w}
 V^{t_{n},z_{n},\gamma_{n}}_{\theta_{n}}&\ge& w(\theta_{n},\hat X^{t_{n},z_{n},\gamma_{n}}_{\theta_{n}}) + \J( \hat X^{t_{n},z_{n},\gamma_{n}}_{\theta} , Y^{t_{n},z_{n},\gamma_{n}}_{\theta_{n}}),
 \ee
  where $z_{n}:=(\xrm(x_{n},y_{n}),y_{n},v_{n}+\J(x_{n},y_{n}))$ and $\theta_{n}$ is the first exit time after $t_{n}$ of $(\cdot,\hat X^{t_{n},z_{n},\gamma_{n}})$ from $B$ (note that $\hat X^{t_{n},z_{n},\gamma_{n}}_{t_{n}}=\xrm(\xrm(x_{n},y_{n}),-y_{n})=x_{n}$). In the following, we use the simplified notations $X^{n},\hat X^{n},$ $V^{n}$ and $Y^{n}$ for the corresponding quantities indexed by  $(t_{n},z_{n},\gamma_{n})$. Since $(t_{o},x_{o})$ reaches a strict minimum $w_{*}-\vp$, this implies
 \be\label{eq: sur sol eq V ge vp}
 V^{n}_{\theta_{n}}&\ge& \vp(\theta_{n},\hat X^{n}_{\theta_{n}}) + \J( \hat X^{n}_{\theta} , Y^{n}_{\theta_{n}})+\iota,
 \ee
 for some $\iota>0$. Let $\check Y^{n}$ be as in Lemma \ref{lem : ito V-w} and observe that 
 \be\label{eq: che Yn - Yn explicit pour sur sol}
 \check Y^{n}-Y^{n}=\frac{\hat X^{n}+\partial_{x}\vp(\cdot,\hat X^{n}) f(\hat X^{n}) -\xrm(\hat X^{n},Y^{n})}{ f(\xrm(\hat X^{n},Y^{n}))}. 
 \ee
 Set 
 $$
 \chi^{n}:= \frac{ (\mu- f'f (a^{n}_{s})^{2}/2)(X^{n})}{\sigma(X^{n})}+ \frac{ \hat F\vp(\cdot,\hat X^{n},Y^{n})}{(Y^{n}-\check Y^{n})\sigma(X^{n})}  \1_{|Y^{n}-\check Y^{n}|\ge \eps}
 $$
 and consider the measure $\P^{n}$ defined by
 $$
 \frac{d\P_{n}}{d\P}=M^{n}_{\theta_{n}} \mbox{ where } M^{n}=1-\int_{t_{n}}^{\cdot\wedge \theta_{n}} M^{n}_{ {s}}\chi^{n}_{s}dW_{s}.
 $$
 Then, it follows from \reff{eq: sur sol eq V ge vp}, Lemma \ref{lem : ito V-w}, \reff{eq: sur sol eq contra t<T} and \reff{eq: che Yn - Yn explicit pour sur sol} that
\begin{align*}
 \iota&\le \E^{\P_{n}}[V^{n}_{\theta_{n}}-(\vp+\J)(\theta_{n},\hat X^{n}_{\theta_{n}}, Y^{n}_{\theta_{n}}) ]
 \\
 &\le v_{n}+\J(x_{n},y_{n})-(\vp+\J)(t_{n},\xrm(\xrm(x_{n},y_{n}),-y_{n}),y_{n})\\
 &= v_{n}-\vp(t_{n},x_{n}).
\end{align*}
  The right-hand side goes to $0$, which is the required contradiction.

{b.} We now explain how to modify the above proof for the case $t_{o}=T$.  After possibly replacing $(t,x)\mapsto \vp(t,x)$ by $(t,x)\mapsto \vp(t,x)-\sqrt{T-t}$, we can assume that $\partial_{t}\vp(t,x)\to \infty$ as $t\to T$, uniformly in $x$ on each compact set.   Then \reff{eq: sur sol eq contra t<T} still holds for $B$ of the form $[T-\eta,T)\x B(x_{o})$ in which $B(x_{o})$ is an open ball around $x_{o}$ and $\eta>0$ {small}. Assume that $\vp(T,x_{o})<G(x_{o})$. Then, after possibly changing $B(x_{o})$, we have $\vp(T,\cdot)\le G-\iota_{1}$ on $B(x_{o})$, for some $\iota_{1}>0$. Then, with the  notations of a., we deduce from \reff{eq: w(T,x)}-\reff{eq: proof sur sol Y ge w} that
\b*
 V^{n}_{\theta_{n}}&\ge& \vp(\theta_{n},\hat X^{n}_{\theta_{n}}) + \J( \hat X^{n}_{\theta} , Y^{n}_{\theta_{n}})+\iota_{1}\wedge \iota_{2},
 \e*
 in which $\iota_{2}:=\min\{(w_{*}-\vp)(t,x) : (t,x)\in [t_{o}-\eta,T)\x \partial B(x_{o})\}>0$ and $\theta_{n}$ is now the minimum between $T$ and the first time after $t_{n}$ at which $\hat X^{n}$ exists $B(x_{o})$.
 The contradiction is then deduced from the same arguments as above.\ep 
\vs2

\noindent {\bf Subsolution property.} We now turn to the subsolution property. Again the proof is close to \cite{bouchard2009stochastic}, except that we have to account for the specific form of  the dynamic programming principle stated in Proposition \ref{prop: DPP}(ii).  Let $\vp$ be a $C^{\infty}_{b}$ function, and  $(t_{o},x_{o})\in [0,T]\x \X$ be a strict (local)  maximum point of $w^{*}-\vp$ such that $(w^{*}-\vp)(t_{o},x_{o})=0$. By \cite[Lemma 4.2]{barlessolutions}, we can find a sequence $(k_{n},t_{n},x_{n})_{n\ge 1}$ such that $k_{n}\to \infty$, $(t_{n},x_{n})$ is a local maximum point of $w_{k_{n}}^{*}-\vp$ and   $(t_{n},x_{n},w_{k_{n}}(t_{n},x_{n}))\to (t_{o},x_{o},w^{*}(t_{o},x_{o}))$.

{a.} As above, we first assume that $t_{o}<T$.    Set   $\vp_{n}(t,x):=\vp(t,x)+|t-t_{n}|^{2}+|x-x_{n}|^{4}$ and assume that $F\vp(t_{o},x_{o})>0$. Then,   $  F\vp_{n}>0$ on a open neighborhood $B$ of $(t_{o},x_{o})$  which contains $(t_{n},x_{n})$, for all $n$ large enough. Since we are going to localize the dynamics, we  can modify $\vp_{n},  \sigma,  \mu$ and $f$ in such a way that they are identically equal to $0$ outside a compact  $A\supset B$. It then follows from Remark \ref{rem : verification} a.~that, after possibly changing  $n\ge 1$, we can  find  $(b^{n},a^{n})\in \Ac_{k_{n}}$ such that the following admits a  strong solution:
\b*
X^{n}&=&x_{n}+\I(x_{n},\hat y[\vp_{n}](t_{n},x_{n}))+\int_{t_{n}}^{\cdot} \sigma(X^{n}_{s}) dW_{s} + \int_{t_{n}}^{\cdot} f(X^{n}_{s}) d Y^{n,c}_{s}\\
&&+\;\int_{t_{n}}^{\cdot} (\mu(X_{s})+a^{n}_{s}(\sigma f')(X^{n}_{s}) )ds
\\
Y^{n}&=&\hat y[\vp_{n}](t_{n},x_{n})+\int_{t_{n}}^{\cdot} b^{n}_{s} ds + \int_{t_{n}}^{\cdot} a^{n}_{s} dW_{s}\\
&=&\xrm^{-1}(\hat X^{n},\hat X^{n}+(f\partial_{x}\vp_{n})(\cdot,\hat X^{n}))\\
\hat X^{n}&:=& \xrm(X^{n},-Y^{n})\\
V^{n}&=&v_{n}+\J(x_{n},\hat y[\vp_{n}](t_{n},x_{n}) )+ \int_{t_{n}}^{\cdot} Y^{n}_{s}dX^{n,c}_{s}+\frac12 \int_{{t_{n}}}^{\cdot} (a^{n}_{s})^{2} f(X^{n}_{s}) ds.
\e*
In the above, we have set $v_{n}:=w_{k_{n}}(t_{n},x_{n})-n^{-1}$.  Observe that the construction of $Y^{n}$ ensures that it coincides with the corresponding process $\check Y^{n}$ of Lemma \ref{lem : ito V-w}. 
  Also note  that $\hat X^{n}_{t_{n}}=\xrm(\xrm(x_{n},y_{n}),$ $-y_{n})$ $=$ $x_{n}$, and let $\theta_{n}$ be the first time after $t_{n}$ at which $(\cdot,  \hat X^{n} )$ exists $B$.   By applying It\^{o}'s Lemma, using Lemma \ref{lem : ito V-w} and the fact that $  F\vp_{n}\ge 0$ on  $B$, we obtain
$$
V^{n}_{\theta_{n}}\ge (\vp_{n}+\J)(\theta_{n},\hat X^{n}_{\theta_{n}},Y^{n}_{\theta_{n}})+v_{n}-\vp_{n}(t_{n},x_{n}).
$$
Let $2\eps:=\min\{|t-t_{o}|^{2}+|x-x_{o}|^{4},\; (t,x) \in \partial B\}$.  For $n$ large enough, the above implies
$$
V^{n}_{\theta_{n}}\ge (w_{k_{n-1}}+\J)(\theta_{n},\hat X^{n}_{\theta_{n}},Y^{n}_{\theta_{n}})+\eps+\iota_{n},
$$
where  $\iota_{n}:=(\vp_{n}-w_{k_{n-1}})(t_{{n-1}},x_{n-1})+v_{n}-\vp_{n}(t_{n},x_{n})$ converges to $0$. Hence, we can find $n$ such that
$$
V^{n}_{\theta_{n}}> (w_{k_{n-1}}+\J)(\theta_{n},\hat X^{n}_{\theta_{n}},Y^{n}_{\theta_{n}}).
$$
Now observe that we can change the subsequence $(k_{n})_{n\ge 1}$ in such a way that $k_{n}\ge 2k_{n-1}+2$. Then,  $v_{n}=w_{k_{n}}(t_{n},x_{n})-n^{-1}< w_{2k_{n-1}+2}(t_{n},x_{n})$, which leads to a contradiction to Proposition \ref{prop: DPP}(ii).

b. It remains to consider the case $t_{o}=T$. As in Step 1., we only explain how to modify the argument used above. Let $(v_{n},k_{n},t_{n},x_{n})$ be as in a. We now set
$\vp_{n}(t,x):=\vp(t,x)+\sqrt{T-t}+|x-x_{n}|^{4}$. Since $\partial_{t} \vp_{n}(t,x)\to -\infty$ as $t\to T$, we can find $n$ large enough so that
$F\vp_{n}\ge 0$ on $[t_{n},T)\x B(x_{o})$ in which $B(x_{o})$ is an open ball around $x_{o}$.  Assume that $\vp(T,x_{o})>G(x_{o})+\eta$ for some $\eta>0$. Then, after possibly changing  $B(x_{o})$, we can assume that $\vp_{n}(T,\cdot)\ge G+\eta$ on $B(x_{o})$. We now use the same construction as in a.~but with $\theta_{n}$ defined as the minimum between $T$ and the first time where $\hat X^{n}$ exists $B(x_{o})$. We obtain
$$
V^{n}_{\theta_{n}}\ge (\vp_{n}+\J)(\theta_{n},\hat X^{n}_{\theta_{n}},Y^{n}_{\theta_{n}})+v_{n}-\vp_{n}(t_{n},x_{n}).
$$
Let $2\eps:=\min\{|x-x_{o}|^{4},\; x \in \partial B(x_{o})\}$. For $n$ large enough, the above implies
\b*
V^{n}_{\theta_{n}}\ge w_{k_{n-1}}(\theta_{n},\hat X^{n}_{\theta_{n}})\1_{\theta_{n}<T}+G(\hat X^{n}_{\theta_{n}})\1_{\theta_{n}=T}+\J(\hat X^{n}_{\theta_{n}},Y^{n}_{\theta_{n}})+\eps\wedge \eta+\iota_{n},
\e*
where  $\iota_{n}$ converges to $0$. By \reff{eq: w(T,x)} and {\rm (H3)},
\b*
V^{n}_{\theta_{n}}> w_{k_{n-1}}(\theta_{n},\hat X^{n}_{\theta_{n}})+\J(\hat X^{n}_{\theta_{n}},Y^{n}_{\theta_{n}}),
\e*
for $n$ large enough. We conclude as in a.\ep

%%%%%%%%%%%%%%%%%%%%%
\subsubsection{Comparison}\label{sec: comparison}

In all this section, we work under the additional condition 
\be\label{eq: inf f >0}
\inf f >0.
\ee
Direct computations (use \reff{eq: PDE} and Proposition \ref{prop: xrm and J}) show that $\hat F \vp$ is of the form 
\be\label{eq: forme generale edp}
\hat F \vp=-\partial_{t}\vp-B(\cdot,f\partial_{x}\vp)\partial_{x}\vp-\frac 12A^{2}(\cdot,f\partial_{x}\vp)\partial_{xx}\vp-L(\cdot,f\partial_{x}\vp)
\ee
where  $A,B$ and $L$ $:$ $(t,x,p)\in [0,T]\x \R\x \R \to\R$  are Lipschitz continuous functions. 

Let $\Phi$ be a solution of the ordinary differential equation 
\be\label{eq: def PHI}
\Phi'(t)=f(\Phi(t)), \;t\in \R.
\ee
Then, $\Phi$ is a bijection on $\R$ (as $f$ is Lipschitz and  $1/f$ {is} bounded) and the following is an immediate consequence of the definition of viscosity solutions. 
\begin{Lemma}\label{lem: change variable pde} Let $v$ be a supersolution (resp. subsolution) of \reff{eq: PDE up to T}. Fix $\rho>0$. Then, $\tilde v$ defined by 
\b*
\tilde v(t,x)=e^{\rho t} v(t,\Phi(x)),
\e*
  is a supersolution (resp. subsolution)  of 
\be
0&=&\rho \vp -\partial_{t}\vp-\left[B(\Phi,e^{-\rho t}\partial_{x}\vp)/f(\Phi)-\frac12A^{2}(\Phi,e^{-\rho t}\partial_{x}\vp)f'(\Phi)/f(\Phi)^{2} \right]\partial_{x}\vp \nonumber\\
&&-\frac 12A^{2}(\Phi,e^{-\rho t}\partial_{x}\vp)\partial_{xx}\vp/f(\Phi)^{2}-e^{\rho t}L(\Phi,e^{-\rho t}\partial_{x}\vp)\label{eq: edp by change variable}
\ee
with the terminal condition 
\be\label{eq: edp bord by change variable}
\vp(T,\cdot)=e^{\rho T} G(\Phi).
\ee

\end{Lemma}

To prove that comparison holds for \reff{eq: PDE up to T}, it suffices to prove that it holds for \reff{eq: edp by change variable}-\reff{eq: edp bord by change variable}. For the latter, this is a consequence of the following result. It is rather standard but we provide the complete proof by lack of a precise reference. 

\begin{Theorem}\label{thm:comp}
Let $\Oc$ be an open subset of $\mathbb{R}$,  $u$ (resp.~$v$) be a upper-semicontinuous subsolution (resp.~lower-semicontinuous supersolution) on $[0,T)\x \Oc$ of:
\begin{equation} \label{eq:eq1}
\rho \vp-\partial_{t}\vp-\bar B(\cdot,e^{-\rho t}\partial_{x}\vp)\partial_{x}\vp-\frac 12\bar A^{2}(\cdot,e^{-\rho t}\partial_{x}\vp)\partial_{xx}\vp-e^{\rho t}\bar L(\cdot,e^{-\rho t}\partial_{x}\vp)=0
\end{equation}
where $\rho>0$ is constant, $\bar A,\bar B$ and $\bar L$ $:$ $(t,x,p)\in [0,T]\x \Oc\x \R \to\R$  are Lipschitz continuous functions. Suppose that $u$ and $v$ are bounded and satisfy $u \leq v$ on the parabolic boundary of $[0,T)\x \Oc$, then $u \leq v$ on the closure of  $[0,T]\x \Oc$.
\end{Theorem}

\proof   Suppose to the contrary that
$$\sup_{[0,T]\x \Oc}(u-v)>0,$$
and define, for $n >0$,
\b*
\Theta_{n}:=\sup_{(t,x,y)\in [0,T)\x \Oc^{2}} \left(u(t,x)-v(t,y)-\frac{n}{2}|x-y|^{2} -\frac {1}{2n} |x|^{2} \right).
\e*
Then, there exists $\iota>0$, such that $\Theta_{n}\ge \iota$ for $n$ large enough. Since $u$ and $v$ are bounded and $u\le v$ on the parabolic boundary of the domain, we can find $(t_{n},x_{n},y_{n})\in [0,T)\x \Oc^{2}$ which achieves the above supremum.

As usual, we apply Ishii's Lemma combined with the sub- and super-solution properties of $u$ and $v$, and the Lipschitz continuity of $\bar A,\bar B$ and $\bar L$ to obtain, with the notation $p_{n}:=n(x_{n}-y_{n})$, 
\b*
\rho (u(t_{n},x_{n})- v(t_{n},y_{n}))   
&\leq& [\bar B(x_{n},e^{-\rho t_{n}}(p_n+\frac1n x_{n}))-\bar B(y_{n},e^{-\rho t_{n}}p_n)]p_n\\
&&+ \frac1n x_{n}\bar B(x_{n},e^{-\rho t_{n}}(p_n+\frac1n x_{n}))\\
&&+\frac{3n}{2}[\bar A(x_{n},e^{-\rho t_{n}}(p_n+\frac1n x_{n}))-\bar A(y_{n},e^{-\rho t_{n}}p_n)]^{2}\\
&& + \frac{1}{2n} \bar A^{2}(x_{n},e^{-\rho t_{n}}(p_n+\frac1n x_{n}))\\
&&+ e^{\rho t_{n}}\left(\bar L(x_{n},e^{-\rho t_{n}}(p_n+\frac1n x_{n}))-\bar L(y_{n},e^{-\rho t_{n}}p_n)\right)\\
&\le& C\left(n(x_{n}-y_{n})^{2}+|x_{n}-y_{n}|+ \frac1n x_{n}^{2} +\frac1n \right)
\e*
for some constant $C$ which does not depend on $n$. 
In view of Lemma \ref{lem: pena comp} below, and since $\rho>0$ and  $u(t_{n},x_{n})- v(t_{n},y_{n})\ge \Theta_{n}\ge \iota$, the above leads to a contradiction for $n$ large enough.
\ep\\
 
We conclude with the proof of the technical lemma that was used in our arguments above.
\begin{Lemma}\label{lem: pena comp}
Let  $\Psi$ be a bounded upper-semicontinuous function on $[0,T]\x  \R^{2}$, and $\Psi_{i}$, $i=1,2$, be two non-negative lower-semicontinuous functions on $\R$ such that  $\{\Psi_{1}=0\}=\{0\}$.
For $n>0$, set 
$$\Theta_{n} := \sup_{(t,x,y)\in [0,T]\x\R^{2}}\left(\Psi(t,x,y)-n\Psi_{1}(x-y)-\frac 1n \Psi_{2}(x)\right)
$$
and assume that there exists  $(\hat t_{n},\hat{x}_{n},\hat{y}_{n}) \in [0,T]\x\R^{2}$  such that:
$$ \Theta_{n}=\Psi(\hat{t}_{n},\hat{x}_{n},\hat{y}_{n} )-n\Psi_{1}(\hat{x}_{n}-\hat{y}_{n} )-\frac 1n \Psi_{2}(\hat{x}_{n}).
$$
Then, after possibly passing to a subsequence, 
\begin{enumerate}[{\rm (i)}]
\item
$\lim\limits_{n \to \infty}n\Psi_{1}(\hat{x}_{n}-\hat{y}_{n} )=0$ and $\lim\limits_{n \to \infty}\frac 1n\Psi_{2}(\hat{x}_{n})=0$.
\item $\lim\limits_{n \to \infty}\Theta_{n}=\sup\limits_{(t,x)\in [0,T]\x\mathcal{O}}\Psi(t,x,x)$.
\end{enumerate}
\end{Lemma}
\proof For later use, set $\bar \R:=\R\cup\{-\infty\}\cup\{\infty\}$ and note that we can extend $\Psi$ as a bounded upper-semicontinuous function on $[0,T]\x\bar \R^{2}$. 
Set  $M := \sup\limits_{(t,x)\in [0,T]\x\R}\Psi(t,x,x)$, and select a sequence $(t_{n},x_{n})_{n\ge 1}$ such that
\b*
\lim_{n \to \infty}\Psi(t_{n},x_{n},x_{n})=M \hspace{3mm}\text{ and }\hspace{3mm}
\lim_{n \to \infty}\frac 1n \Psi_{2}(x_{n})=0.
\e*
Let $C$ be a upper-bound for $\Psi$. Then, 
\b*
C-n\Psi_{1}(\hat{x}_{n}-\hat{y}_{n})-\frac 1n \Psi_{2}(\hat{x}_{n})&\geq& \Psi(\hat t_{n},\hat{x}_{n},\hat{y}_{n})-n\Psi_{1}(\hat{x}_{n}-\hat{y}_{n})-\frac 1n \Psi_{2}(\hat{x}_{n})\\
&\geq& \Psi(t_{n},x_{n},x_{n})-\frac 1n \Psi_{2}(x_{n})\\
&\geq& M-\eps_{n}
\e*
where $\epsilon_{n}\to 0$. Since $\Psi_{1}$ and $\Psi_{2}$ are non-negative,  letting $n \to \infty$ in the above inequality leads to
$$\lim_{n \to \infty}\Psi_{1}(\hat{x}_{n}-\hat{y}_{n})=0$$
which implies $\lim_{n \to \infty}(\hat{x}_{n}-\hat{y}_{n})=0$ by the assumption $\{\Psi_{1}=0\}=\{0\}$.

After possibly passing to a subsequence, we can then assume that  $\lim_{n\to \infty} \hat{x}_{n}=\lim_{n\to \infty} \hat{y}_{n}=\hat{x}\in \bar \R $ and that $\lim_{n\to \infty} \hat{t}_{n}=\hat t\in [0,T]$. Since $\Psi$ is upper semi-continuous, the above leads to 
\b*
& &M-\liminf_{n\to \infty}\left(n\Psi_{1}(\hat{x}_{n}-\hat{y}_{n})+\frac 1n \Psi_{2}(\hat{x}_{n})\right) \\
&\geq& \Psi(\hat t,\hat{x},\hat{x})-\liminf_{n\to \infty}\left(n\Psi_{1}(\hat{x}_{n}-\hat{y}_{n})-\frac 1n \Psi_{2}(\hat{x}_{n}) \right)\\
&\geq& \limsup_{n\to \infty}\left(\Psi(\hat t_{n},\hat{x}_{n},\hat{y}_{n})-n\Psi_{1}(\hat{x}_{n}-\hat{y}_{n})-\frac 1n \Psi_{2}(\hat{x}_{n}) \right) \\
&\geq& M,
\e*
and our claim follows.  \ep

\begin{Remark} It follows from the above that, whenever they are bounded, e.g. if $G$ is bounded, then $w_{*}\ge w^{*}$. Since by construction $w_{*}\le w \le  w^{*}$, the three functions are equal to the unique bounded viscosity solution of \reff{eq: PDE up to T}.
\end{Remark}

%%%%%%%%%%%%%%%%%%
\subsubsection{Smoothness}

We conclude here the proof of Theorem \ref{thm : pde} by showing that existence of a smooth solution holds when 
\be\label{eq: hyp f G}
\mbox{$\inf f>0$, $G$ is bounded and  $C^{2}$ with $G,G',G^{''}$ H\"older continuous.}
\ee
Note that the assumptions $\inf f>0$ and {\rm (H1)}  imply that $\Phi^{-1}$ is $C^{2}${, recall \reff{eq: def PHI}}. Hence, by the same arguments as in Section \ref{sec: comparison}, existence of a $C^{1,2}([0,T)\x \R)\cap C^{0}([0,T]\x \R)$ solution to \reff{eq: edp by change variable}-\reff{eq: edp bord by change variable} implies the existence  of a $C^{1,2}([0,T)\x \R)\cap C^{0}([0,T]\x \R)$ solution to \reff{eq: PDE up to T}. As for  \reff{eq: edp by change variable}-\reff{eq: edp bord by change variable}, this is a consequence of  \cite[Thm 14.24]{lieberman1996second}, under {\rm (H1)} and \reff{eq: hyp f G}. 
 
 It remains to show that the solution can be taken bounded, then the comparison result of Section \ref{sec: comparison} will imply that $w$ is this solution. Again, it suffices to work with \reff{eq: edp by change variable}-\reff{eq: edp bord by change variable}.   Let $\vp$ be a $C^{1,2}([0,T)\x \R)\cap C^{0}([0,T]\x \R)$ solution of  \reff{eq: edp by change variable}-\reff{eq: edp bord by change variable}. 
 Let $S^{t,x}$ be defined by 
 $$
 S^{t,x}_{s}=x +\int_{t}^{s}\mu_{S}(s,S^{t,x}_{s}) ds +\int_{t}^{s}\sigma_{S}(s,S^{t,x}_{s})  dW_{s} ,\; s\ge t,
 $$
 where 
 \b*
 \mu_{S}&:=& B(\Phi,e^{-\rho t}\partial_{x}\vp)/f(\Phi)-\frac12A^{2}(\Phi,e^{-\rho t}\partial_{x}\vp)f'(\Phi)/f(\Phi)^{2}\\
 \sigma_{S}&:=& A(\Phi,e^{-\rho t}\partial_{x}\vp)/f(\Phi).
 \e*
Note that the coefficients of the sde may only be locally Lipschitz. However, they are bounded (recall {\rm (H1)} and \reff{eq: hyp f G}), which is enough to define a solution by a standard localization procedure. 
Since $\sigma_{S}$ is bounded, It\^{o}'s Lemma implies that 
 $$
 \vp(t,x)e^{-\rho t}=\Esp{G(\Phi(S^{t,x}_{T}))+\int_{t}^{T}  L(\Phi(X^{t,x}_{s}),e^{-\rho s}\partial_{x}\vp(s,X^{t,x}_{s}))ds}.
 $$
 Since $G$ and $L$  are bounded, by {\rm (H1)} and \reff{eq: hyp f G},  $\vp$ is bounded as well.\ep

\begin{Remark} We refer to \cite{ladyzhenska1988linear} for conditions under which additional smoothness of the solution can be proven.
\end{Remark}

%%%%%%%%%%%%%%%%%%%%
\appendix
\section{Appendix}
 
We report here the measurability property that was used in the course of Proposition \ref{prop: DPP}.

In the following, $\Ac_{k}$ is viewed as a closed subset of the Polish space  $\Lb^{\lambda}_{2}$ endowed with the usual strong norm topology $\|\cdot\|_{\Lb^{\lambda}_{2}}$.

We consider an element $\nu \in \Uc_{k}$
as  a measurable map $\omega \in \Omega \mapsto \nu(\omega) \in  \Mc_{k}$ where $\Mc_{k}$ denotes the set of non-negative Borel   measures on $\R\x [0,T]$ with total mass less than $k$, endowed with the {topology of weak convergence}. This topology  is generated by the  norm
$$
 \|m\|_{\Mc}:=\sup\{\int_{\R\x [0,T]} \ell(\delta,s)  m(d\delta,ds): \ell\in {\rm Lip}_{1}\},
$$
in which ${\rm Lip}_{1}$ denotes the class of $1$-Lipschitz continuous functions bounded by $1$, see e.g.~\cite[Proposition 7.2.2 and Theorem 8.3.2]{bogachev2007measure}.  Then, $\Uc_{k}$ is  a closed subset of the space  $\Mb_{k,2}$  of  $\Mc_{k}$-valued random variables.
 $\Mb_{k,2}$ is made complete and separable by
the norm
$$
  \|\nu \|_{\Mb_{2}}:=\Esp{\|\nu\|_{\Mc}^{2}}^{\frac12}.
$$
See e.g.~\cite[Chap.~5]{crauel2003random}.
We endow the set of controls $\Gamma_{k}$ with the natural product topology
$$
  \|\gamma\|_{\Lb^{\lambda}_{2}\x\Mb_{2}}:=\|\vartheta\|_{\Lb^{\lambda}_{2}}+\|\nu\|_{\Mb_{2}},\;\mbox{ for } \gamma=(\vartheta,\nu).
$$
As a closed subset of the Polish space $\Lb^{\lambda}_{2}\x \Mb_{k,2}$, $\Gamma_{k}$ is a Borel space, for each $k\ge 1$. See e.g.~\cite[Proposition 7.12]{BertsekasShreve.78}.

The following stability result is proved by using standard estimates. In the following, we use the notation $Z=(X,Y,V)$.
\begin{Proposition}\label{prop : L2 estimates on initial data and control} For each $k\ge1$, there exists a real constant $c_{k}>0$ such that
$$
\|Z^{t_{1},z_{1},\gamma_{1}}_{T}-Z^{t_{2},z_{2},\gamma_{2}}_{T}\|_{\Lb_{2}}\le c_{k}\left(|t_{1}-t_{2}|^{\frac12}+|z_{1}-z_{2}|+\|\gamma_{1}-\gamma_{2}\|_{\Lb^{\lambda}_{2}\x\Mb_{2}} \right),
$$
for all  $(t_{i},z_{i},\gamma_{i})\in \D\x \Gamma_{k}$, $i=1,2$.
\end{Proposition}

A direct consequence is the continuity of $(t,z,\gamma)\in \D\x \Gamma_{k} \mapsto Z^{t,z,\gamma}_{T}$, which is therefore measurable
\begin{Corollary}\label{cor : measurability} For each $k\ge 1$,  the map $(t,z,\gamma)\in \D\x \Gamma_{k} \mapsto Z^{t,z,\gamma}_{T}  \in \Lb_{2}$   is   Borel-measurable.
\end{Corollary}
  %%%%%%%%%%%%%%%%%%%%%%%%%%%%%%%
 \bibliographystyle{plain}

  \end{document}